\newif\ifOneCol

\ifOneCol
	\documentclass[draftcls,onecolumn,11pt]{IEEEtran}
\else
	\documentclass[twocolumn,10pt]{IEEEtran}
\fi

\usepackage[cmex10]{amsmath}
\usepackage{amsfonts}
\usepackage{mathtools}
\usepackage{verbatim}
\usepackage{algpseudocode}
\usepackage{hyperref}

\usepackage{cite}

\ifCLASSINFOpdf
  \usepackage[pdftex]{graphicx}
  \graphicspath{{graphics/}}
  \DeclareGraphicsExtensions{.pdf,.jpeg,.png}
\else
  \usepackage[dvips]{graphicx}
  \graphicspath{{graphics/}}
  \DeclareGraphicsExtensions{.eps}
\fi

\usepackage{array}

\usepackage{mdwmath}
\usepackage{mdwtab}
\DeclareMathOperator{\erf}{erf}

\begin{document}
\bibliographystyle{IEEEtran}
%
\title{Improving Receiver Performance of Diffusive Molecular
Communication with Enzymes}

\author{Adam Noel, \IEEEmembership{Student Member, IEEE}, Karen C. Cheung, and
Robert Schober,
\IEEEmembership{Fellow, IEEE}
\thanks{Manuscript received January 22, 2013; revised October 16, 2013; accepted
December 4, 2013. This
work was presented in part at the 7th International Conference on Bio-Inspired
Models of Network, Information, and Computing Systems (BIONETICS 2012) in
Lugano, Switzerland \cite{RefWorks:631}. This work was supported by the Natural
Sciences and Engineering Research Council of Canada, and a Walter C. Sumner
Memorial Fellowship. Computing resources were provided by WestGrid and
Compute/Calcul Canada.}
\thanks{The authors are with the Department of Electrical and
Computer Engineering, University of British Columbia, Vancouver, BC, Canada,
V6T 1Z4 (email: \{adamn, kcheung, rschober\}@ece.ubc.ca).
R. Schober is also with the Institute for Digital
Communication, Friedrich-Alexander-Universit\"{a}t Erlangen-N\"{u}rnberg (FAU),
Erlangen, Germany
(email: schober@lnt.de).}}


\newcommand{\dbydt}[1]{\frac{d#1}{dt}}
\newcommand{\pbypx}[2]{\frac{\partial #1}{\partial #2}}
\newcommand{\psbypxs}[2]{\frac{\partial^2 #1}{\partial {#2}^2}}
\newcommand{\dbydtc}[1]{\dbydt{\conc{#1}}}
\newcommand{\thev}{\theta_v}
\newcommand{\thevi}[1]{\theta_{v#1}}
\newcommand{\theh}{\theta_h}
\newcommand{\thehi}[1]{\theta_{h#1}}
\newcommand{\x}{x}
\newcommand{\y}{y}
\newcommand{\z}{z}
\newcommand{\rad}[1]{\vec{r}_{#1}}
\newcommand{\radmag}[1]{|\rad{#1}|}

\newcommand{\kth}[1]{k_{#1}}
\newcommand{\km}{K_M}
\newcommand{\vm}{v_{max}}
\newcommand{\conc}[1]{[#1]}
\newcommand{\conco}[1]{[#1]_0}
\newcommand{\C}{C}
\newcommand{\Cx}[1]{C_{#1}}
\newcommand{\CxFun}[3]{C_{#1}(#2,#3)}
\newcommand{\Cobs}{C_{ob}}
\newcommand{\Nobs}{{\Nx{\A}}_{ob}}
\newcommand{\Nobst}[1]{\Nobs\left(#1\right)}
\newcommand{\Nobsn}[1]{\Nobs\left[#1\right]}
\newcommand{\Nobsavgt}{\overline{{\Nx{\A}}_{ob}}(t)}
\newcommand{\Nobsavgn}[1]{\overline{{\Nx{\A}}_{ob}}\left[#1\right]}
\newcommand{\Nobsavg}[1]{\overline{{\Nx{\A}}_{ob}}\left(#1\right)}
\newcommand{\Nobsavgmax}{\overline{{\Nx{\A}}_{max}}}
\newcommand{\Nemit}{{\Nx{\A}}_{em}}
\newcommand{\Cgen}{C_A(r, t)}
\newcommand{\radbind}{r_B}

\newcommand{\M}{M}
\newcommand{\smM}{m}
\newcommand{\A}{A}
\newcommand{\X}{S}
\newcommand{\metre}{\textnormal{m}}
\newcommand{\second}{\textnormal{s}}
\newcommand{\molecule}{\textnormal{molecule}}
\newcommand{\bound}{\textnormal{bound}}
\newcommand{\Dx}[1]{D_{#1}}
\newcommand{\Nx}[1]{N_{#1}}
\newcommand{\Da}{D_\A}
\newcommand{\En}{E}
\newcommand{\en}{e}
\newcommand{\Ne}{N_{\En}}
\newcommand{\De}{D_\En}
\newcommand{\EA}{EA}
\newcommand{\ea}{ea}
\newcommand{\Nint}{N_{\EA}}
\newcommand{\Di}{D_{\EA}}
\newcommand{\Etot}{\En_{Tot}}
\newcommand{\stepl}{r_{rms}}
\newcommand{\AP}{A_P}
\newcommand{\Ri}[1]{R_{#1}}
\newcommand{\ro}{r_0}
\newcommand{\rone}{r_1}
\newcommand{\visc}{\eta}
\newcommand{\bolt}{\kth{B}}
\newcommand{\temp}{T}
\newcommand{\T}{T_B}
\newcommand{\Vobs}{V_{ob}}
\newcommand{\robs}{r_{ob}}
\newcommand{\Ve}{V_{enz}}
\newcommand{\tint}{\delta t}
\newcommand{\tmax}{t_{max}}
\newcommand{\dist}{L}
\newcommand{\DMLSA}{a}
\newcommand{\DMLSt}[1]{t_{#1}^\star}
\newcommand{\DMLSx}{x^\star}
\newcommand{\DMLSy}{y^\star}
\newcommand{\DMLSz}{z^\star}
\newcommand{\DMLSr}{r_{ob}^\star}
\newcommand{\DMLSrad}[1]{\rad{#1}^\star}
\newcommand{\DMLSradmag}[1]{|\DMLSrad{#1}|}
\newcommand{\DMLSC}[1]{\Cx{#1}^\star}
\newcommand{\DMLSc}[1]{\gamma_{#1}}
\newcommand{\DMLSV}{\Vobs^\star}
\newcommand{\DMLSNA}{\overline{{\Nx{\DMLSA}}_{ob}^\star}(t)}
\newcommand{\DMLSNAb}{\overline{{\Nx{\DMLSA}}_{ob}^\star}(\DMLSt{B})}
\newcommand{\DMLSNAmax}{{\overline{{\Nx{\DMLSA}}_{max}^\star}}}
\newcommand{\DMLStmax}[1]{{\DMLSt{#1}}_{,max}}
\newcommand{\DMLSdim}{\mathcal{D}}
\newcommand{\DMLSthreshInt}{\alpha^\star}

\newcommand{\data}[1]{W\left[#1\right]}
\newcommand{\dataObs}[1]{\hat{W}\left[#1\right]}
\newcommand{\thresh}{\xi}
\newcommand{\poissBar}{\Big|_\textnormal{Poiss}}
\newcommand{\gaussBar}{\Big|_\textnormal{Gauss}}
\newcommand{\Pobs}{P_{ob}}
\newcommand{\Pobsx}[1]{P_{ob}\left(#1\right)}
\newcommand{\Pone}{P_1}
\newcommand{\Pzero}{P_0}
\newcommand{\Pe}[1]{P_e\left[#1\right]}
\newcommand{\Peavg}{\overline{P}_e}
\newcommand{\threshInterval}{\alpha}
\newcommand{\threshT}{t_\alpha}

\newcommand{\fof}[1]{f\left(#1\right)}
\newcommand{\floor}[1]{\lfloor#1\rfloor}
\newcommand{\lam}[1]{W\left(#1\right)}
\newcommand{\EXP}[1]{\exp\left(#1\right)}
\newcommand{\ERF}[1]{\erf\left(#1\right)}
\newcommand{\SIN}[1]{\sin\left(#1\right)}
\newcommand{\SINH}[1]{\sinh\left(#1\right)}
\newcommand{\COS}[1]{\cos\left(#1\right)}
\newcommand{\COSH}[1]{\cosh\left(#1\right)}
\newcommand{\Ix}[2]{I_{#1}\!\left(#2\right)}
\newcommand{\Jx}[2]{J_{#1}\!\left(#2\right)}

\newcommand{\B}[1]{B_{#1}}
\newcommand{\w}{w}
\newcommand{\n}{n}

\newcommand{\new}[1]{\textbf{#1}}
\newcommand{\ISI}{ISI}
\newcommand{\PDF}{PDF}

\newtheorem{theorem}{Theorem}

\newcommand{\edit}[2]{
\ifOneCol
	\textbf{#1}
\else
	#1
\fi}

\newcommand{\figEnz}[1]{
	\begin{figure}[#1]
		\centering
		\includegraphics[width=.9\linewidth]{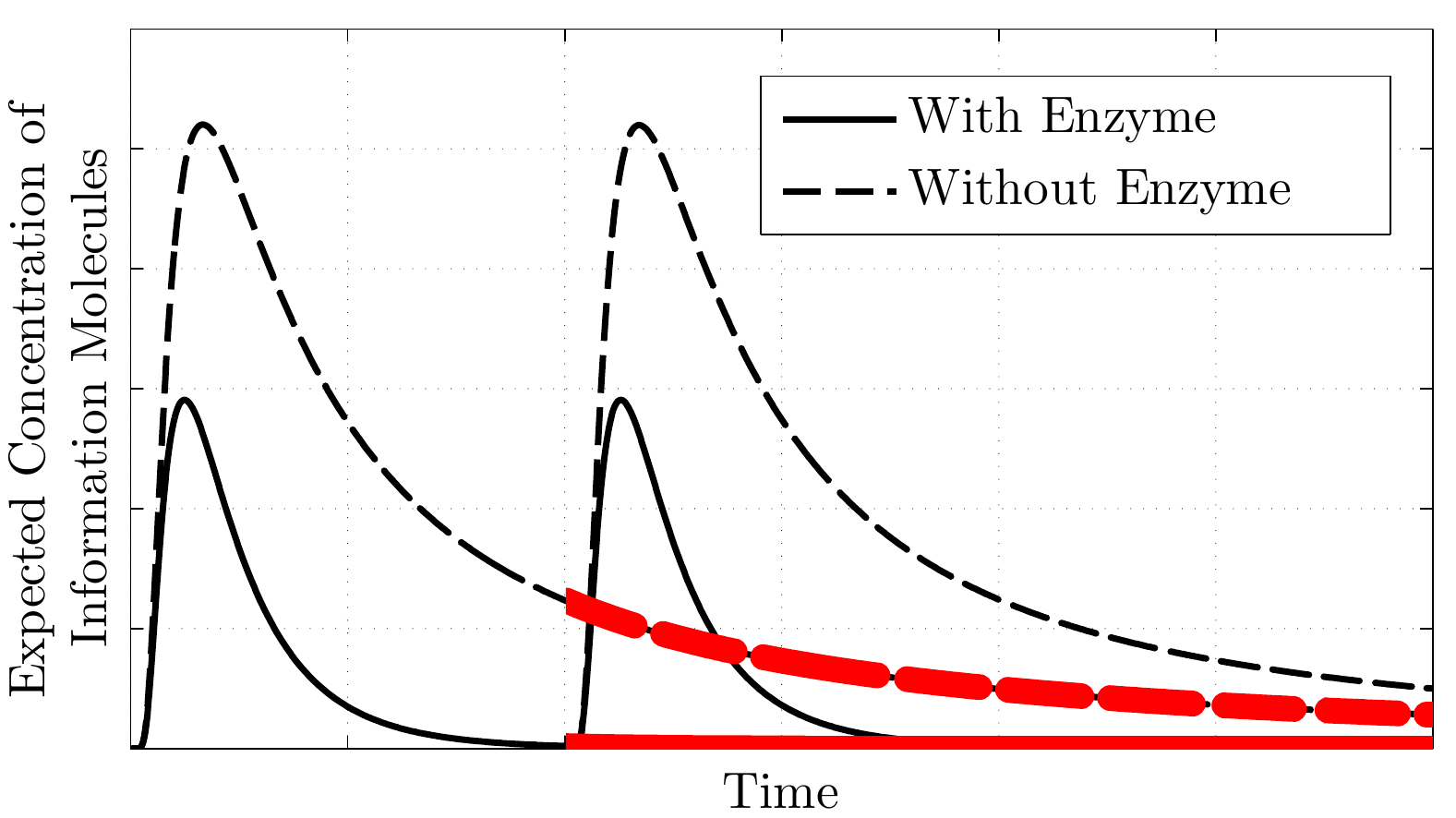}
		\caption{A sample comparison of the expected concentration of information
		molecules at a receiver with and without enzymes present in the propagation
		environment. In each case, the transmitter emits two impulses of molecules.
		The relative quantity of \ISI\, from the first impulse, shown as a thicker
		red line, is much greater without active enzyme molecules.}
		\label{fig_enz}
	\end{figure}
}
\newcommand{\figVenz}[1]{
	\begin{figure}[#1]
		\centering
		\includegraphics[width=.9\linewidth]{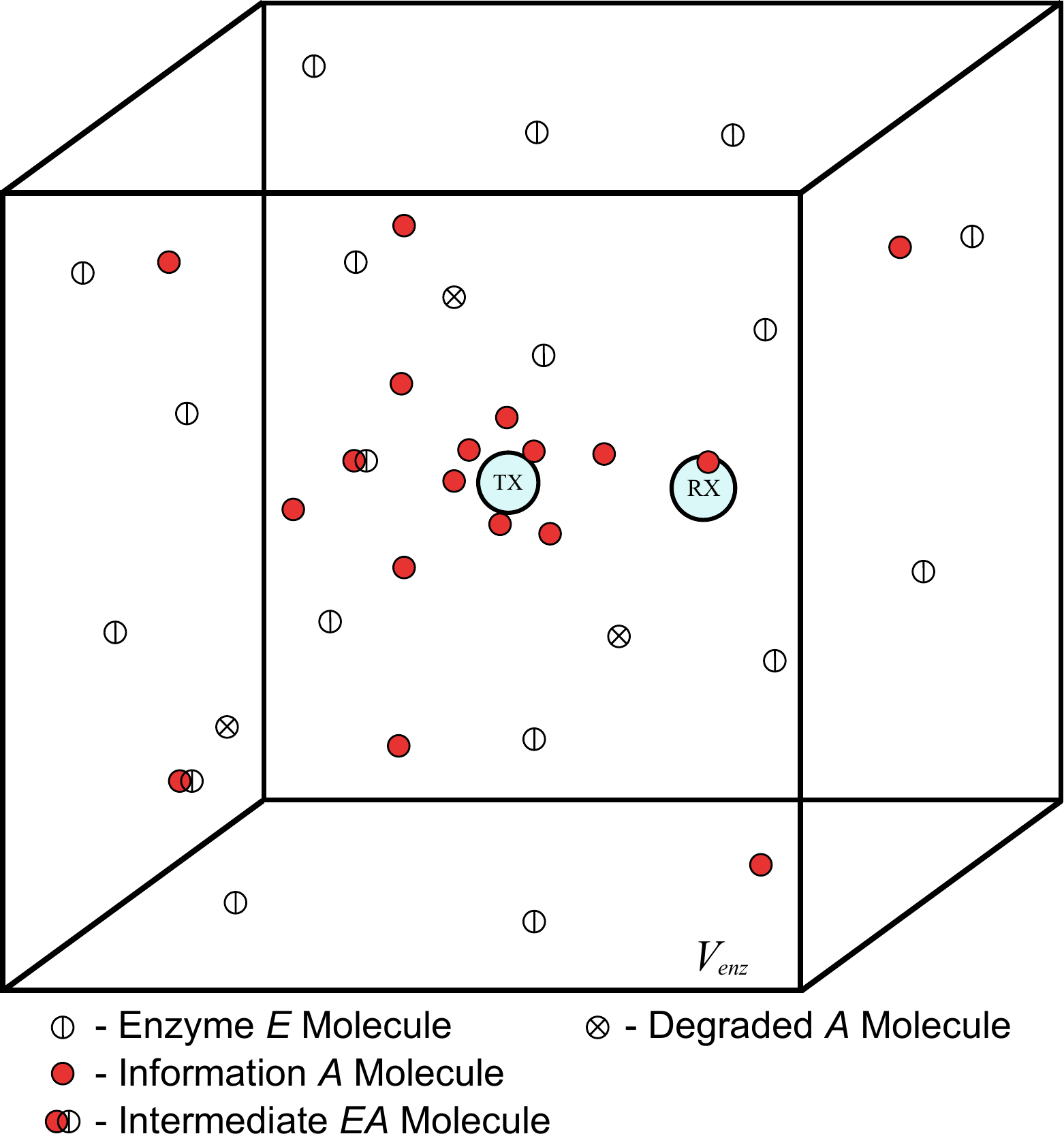}
		\caption{The bounded space $\Ve$ showing a uniform
		distribution of enzyme $\En$ molecules
		(enzymes are shown as circles with vertical
		lines through them). $\Ve$ inhibits the passage of $\En$ so that the
		total concentration of free and bound $\En$ remains constant. Information
		$\A$ molecules (shown as red circles) are emitted by the transmitter and can
		diffuse beyond $\Ve$. Intermediate $\EA$ molecules can form when an $\A$
		molecule binds to an $\En$ molecule. When an intermediate dissociates,
		it can leave the $\A$ molecule degraded
		(shown as a circle with an X through it).}
		\label{venz}
	\end{figure}
}
\newcommand{\tableAccuracy}[1]{
	\begin{table}[#1]\normalsize
	\centering
	\caption{System parameters used for numerical and simulation results.
	The values for $\stepl$ and $\radbind$ are calculated from (\ref{AUG12_25})
	and (\ref{AUG12_26}), respectively.}
	{\renewcommand{\arraystretch}{1.2}
		\begin{tabular}{|c|c|c|c|}
		\hline
		Parameter & System 1 & System 2 & System 3\\ \hline
		$\Ve$ [$\mu\metre^3$] & $1$ & $37$ & $1$ \\ \hline
		$\Nemit$ & $5\times10^3$ & $5\times10^3$ & $2\times10^4$ \\ \hline
		$\Ne$ & $10^5$ & $3.7\times10^6$ & $10^5$ \\ \hline
		$\kth{1}$ [$\frac{\metre^3}{\molecule\cdot\second}$]& 
		$2\times10^{-19}$ & $1.79\times10^{-20}$ & $2\times10^{-19}$ \\ \hline
		$\kth{-1}$ [$\second^{-1}$] & $10^4$ &
		$900$ & $10^4$ \\ \hline
		$\kth{2}$ [$\second^{-1}$] & $10^6$ &
		$9\times10^4$ & $10^6$ \\ \hline
		$\radmag{0}$ [nm] & $300$ & $1000$ & $300$ \\ \hline
		$\radmag{ob}$ [nm] & $45$ & $150$ & $45$ \\ \hline
		$\Ri{\A}$ [nm] & $0.5$ & $0.5$ & $0.5$ \\ \hline
		$\Ri{\En}$ [nm] & $2.5$ & $2.5$ & $2.5$ \\ \hline
		$\Ri{\EA}$ [nm] & $3$ & $3$ & $3$ \\ \hline
		$\Delta t$ [$\mu\second$] & $0.5$ & $5$ & $0.5$ \\ \hline
		$\stepl$ [nm] & $22.9$ & $72.4$ & $22.9$ \\ \hline
		$\radbind$ [nm] & $2.88$ & $2.77$ & $2.88$ \\ \hline 
		\end{tabular}
	}
	\label{table_accuracy}
	\end{table}
}
\newcommand{\figAccuracy}[1]{
	\begin{figure}[#1]
	\centering
	\includegraphics[width=\linewidth]
	{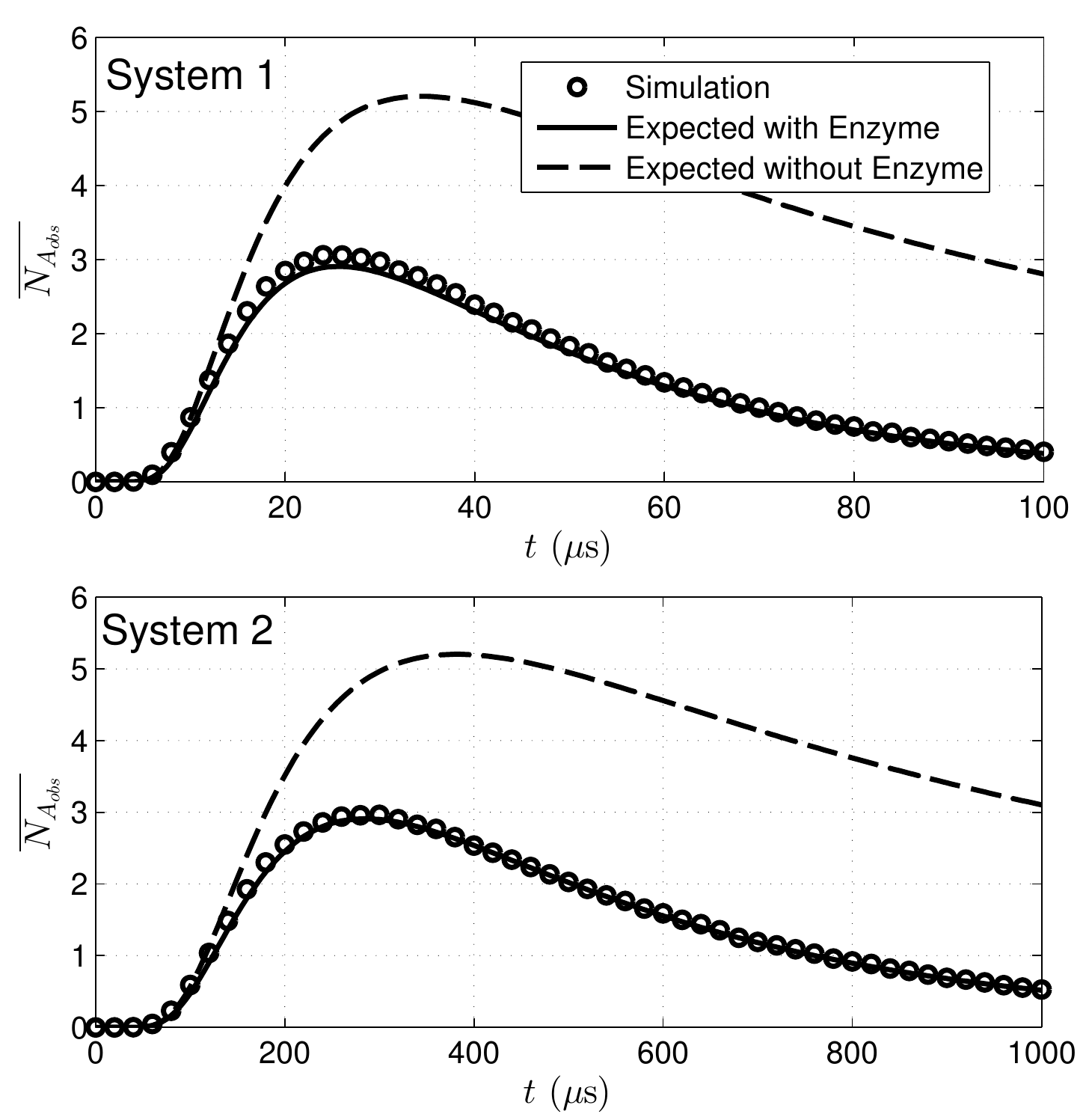}
	\caption{Assessing the accuracy of the lower bound on the expected
	number of observed information molecules for Systems 1 (above) and 2 (below).
	The two systems have the same lower bound on the expected number of
	observed molecules when we account for System 2's longer diffusion time (the
	receiver is placed further away), but this bound is more accurate for System 2.}
	\label{fig_accuracy}
	\end{figure}
}
\newcommand{\figInterval}[1]{
	\begin{figure}[#1]
	\centering
	\includegraphics[width=\linewidth]
	{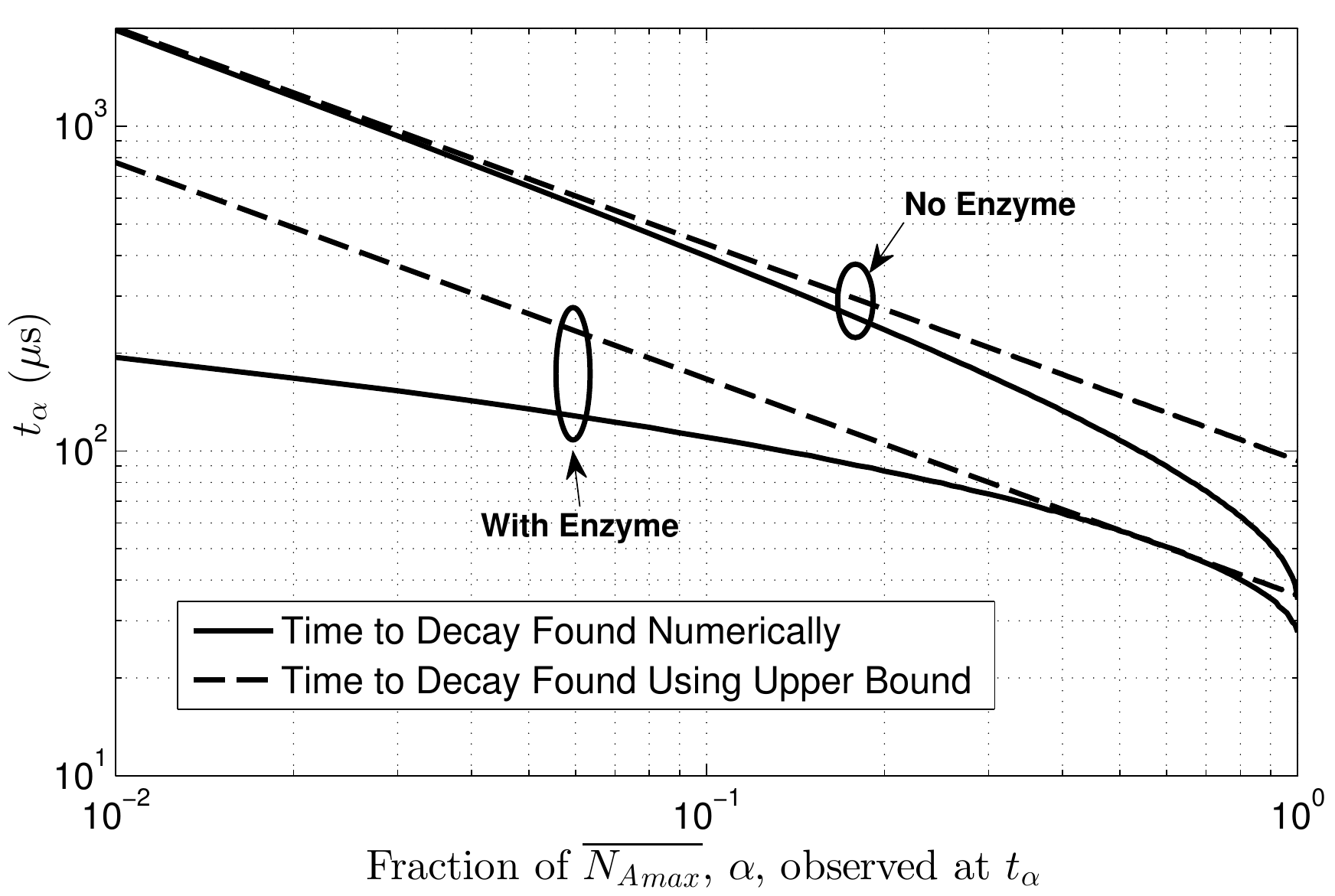}
	\caption{Solving (\ref{JUN12_75}) for System 1 to determine how long it
	would take after a transmitter's single emission for the expected number of
	information molecules to decay to threshold fraction $\threshInterval$. The
	inequality is solved both numerically and by using upper bounds
	(\ref{JUN12_77}) and (\ref{JUN12_80}) for System 1 having enzymes
	present and absent, respectively.}
	\label{fig_interval}
	\end{figure}
}
\newcommand{\figSingle}[1]{
	\begin{figure}[#1]
	\centering
	\includegraphics[width=\linewidth]
	{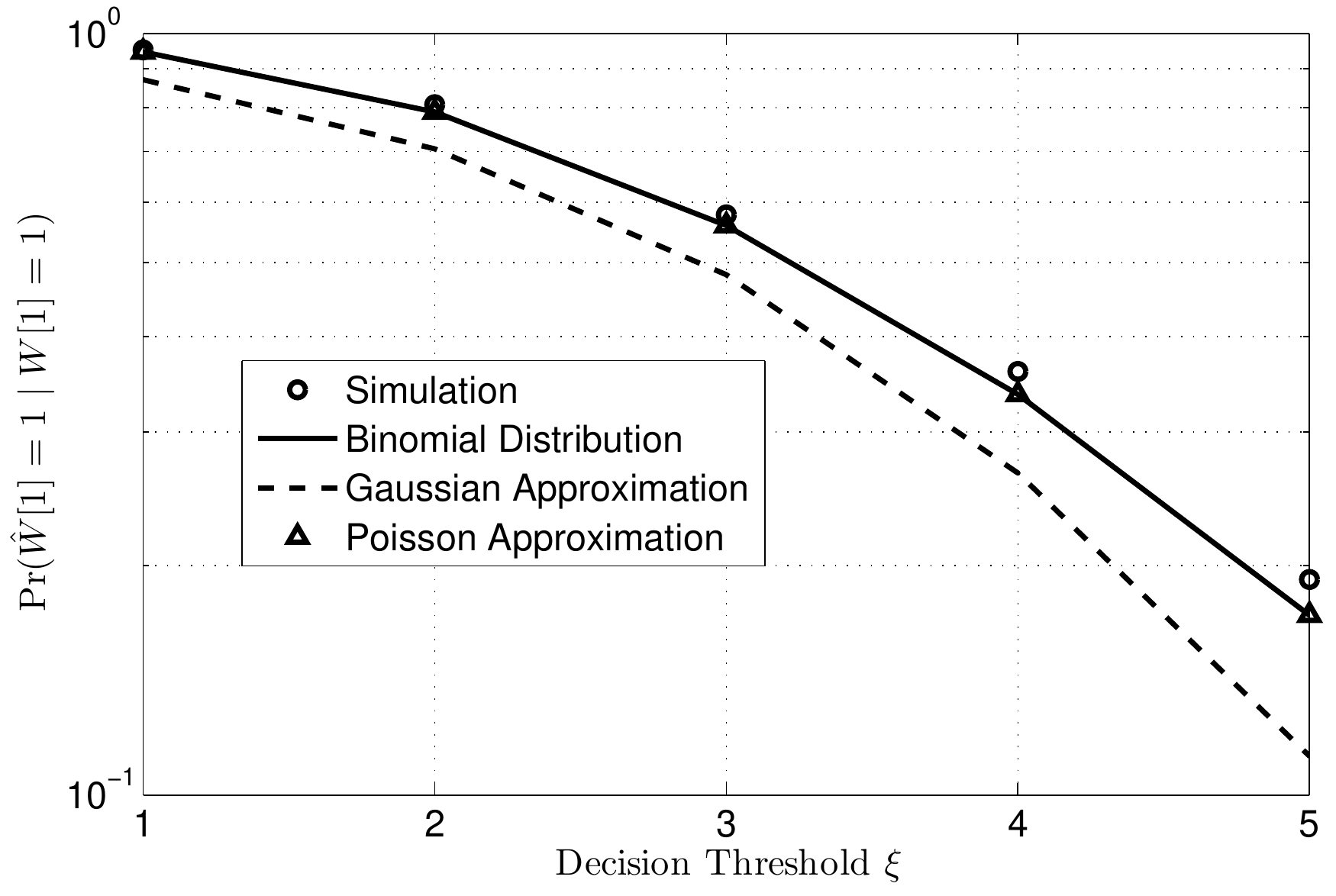}
	\caption{Evaluating the detection probability for the first bit in System 1,
	i.e., $\Pr(\dataObs{1} = 1 | \data{1} = 1)$, as a function of decision
	threshold $\thresh$.}
	\label{fig_single}
	\end{figure}
}
\newcommand{\figKnown}[1]{
	\begin{figure}[#1]
	\centering
	\includegraphics[width=\linewidth]
	{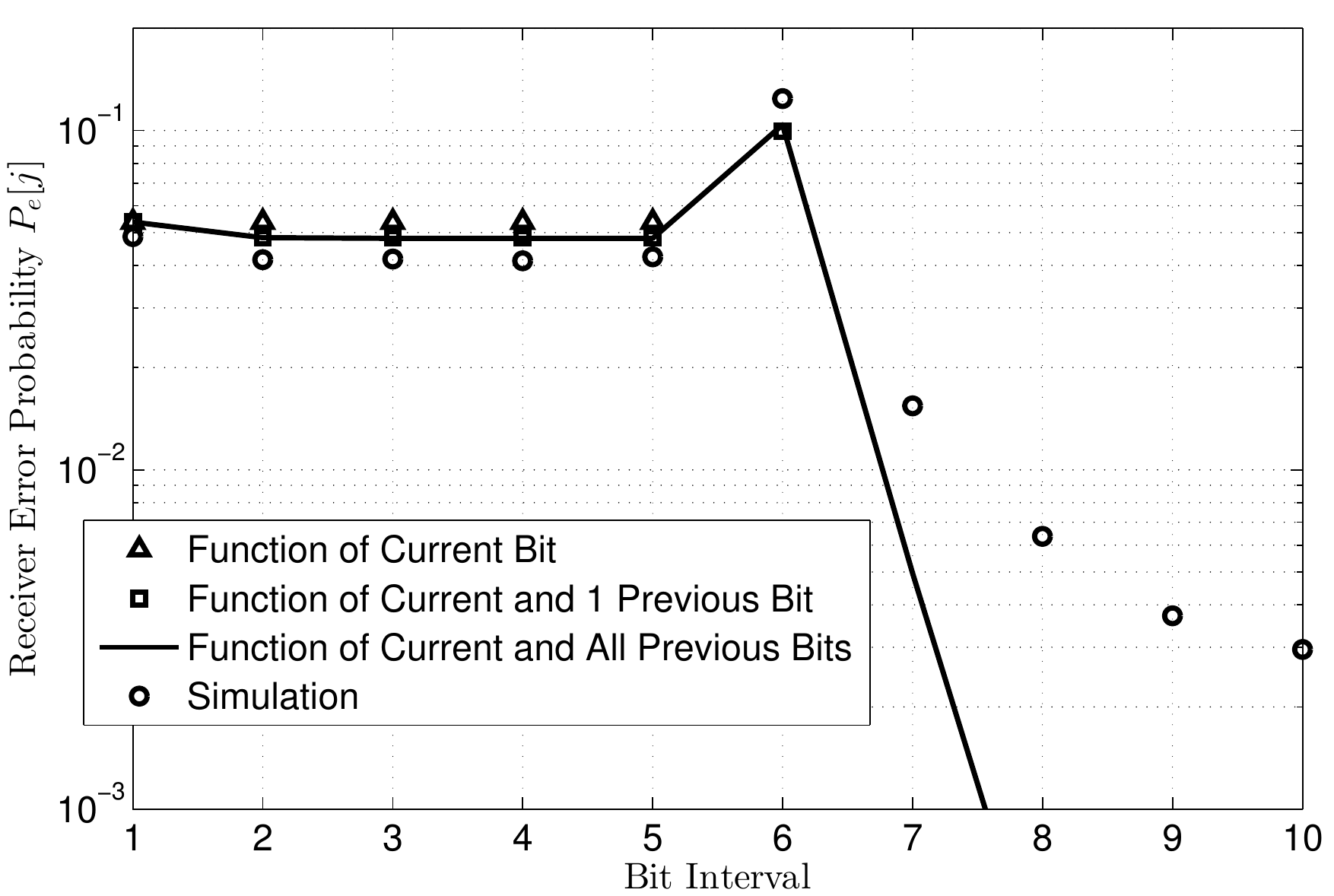}
	\caption{Evaluating the error probability of System 1 over time with
	bit interval $\T =
	120\,\mu\second$ and a known transmission sequence; five $1$s followed by five
	$0$s.}
	\label{fig_known_data}
	\end{figure}
}
\newcommand{\figError}[1]{
	\begin{figure}[#1]
	\centering
	\includegraphics[width=\linewidth]
	{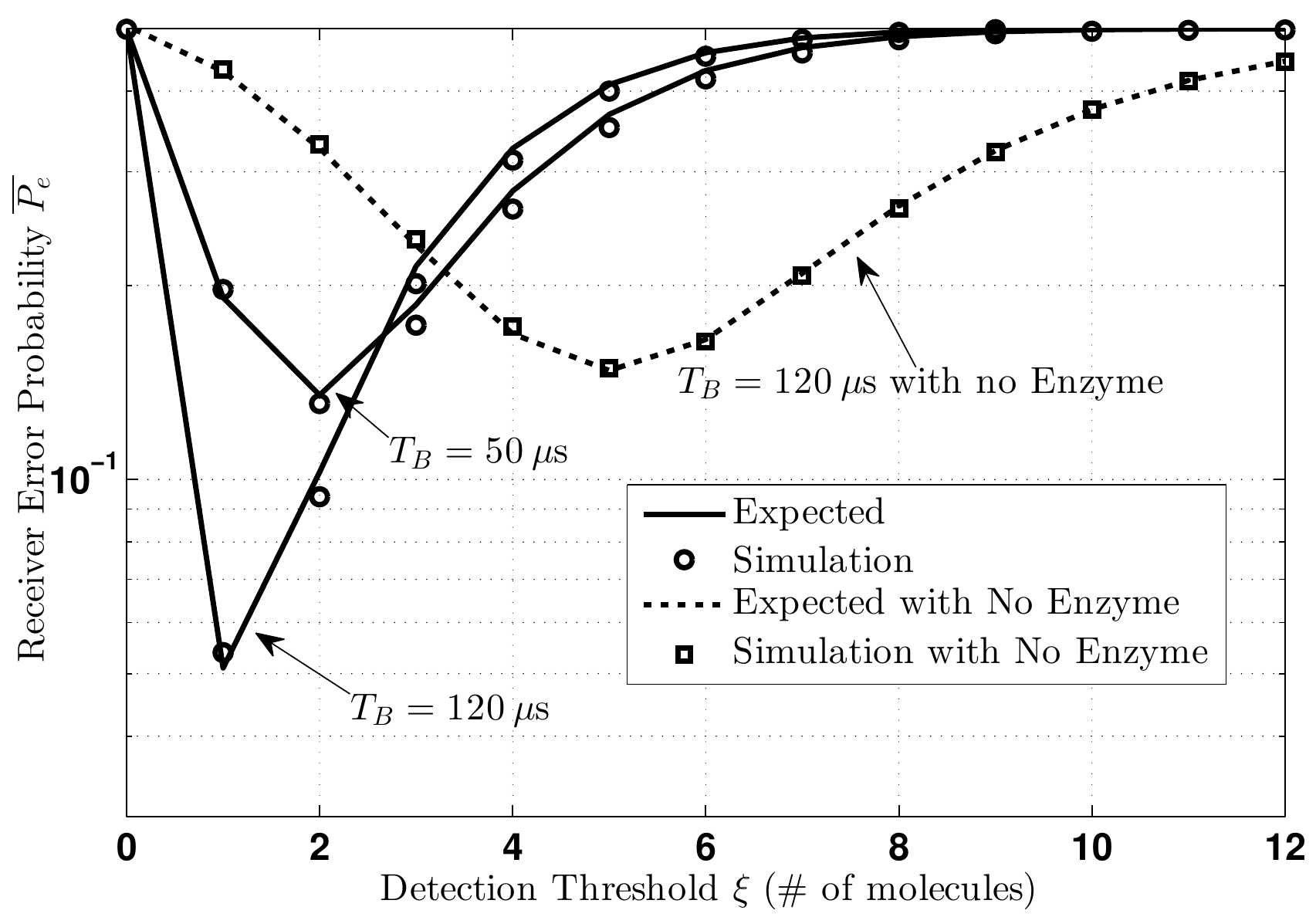}
	\caption{Evaluating the error probability of System 1 as a function of the bit
	decision threshold $\thresh$ at the receiver for bit interval
	$\T=50\,\mu\second$ and
	$\T=120\,\mu\second$ with enzymes and $\T=120\,\mu\second$ without enzymes. The
	transmission is a sequence of $50$ randomly generated bits.}
	\label{fig_error_vs_thresh}
	\end{figure}
}
\newcommand{\figErrorTwo}[1]{
	\begin{figure}[#1]
	\centering
	\includegraphics[width=\linewidth]
	{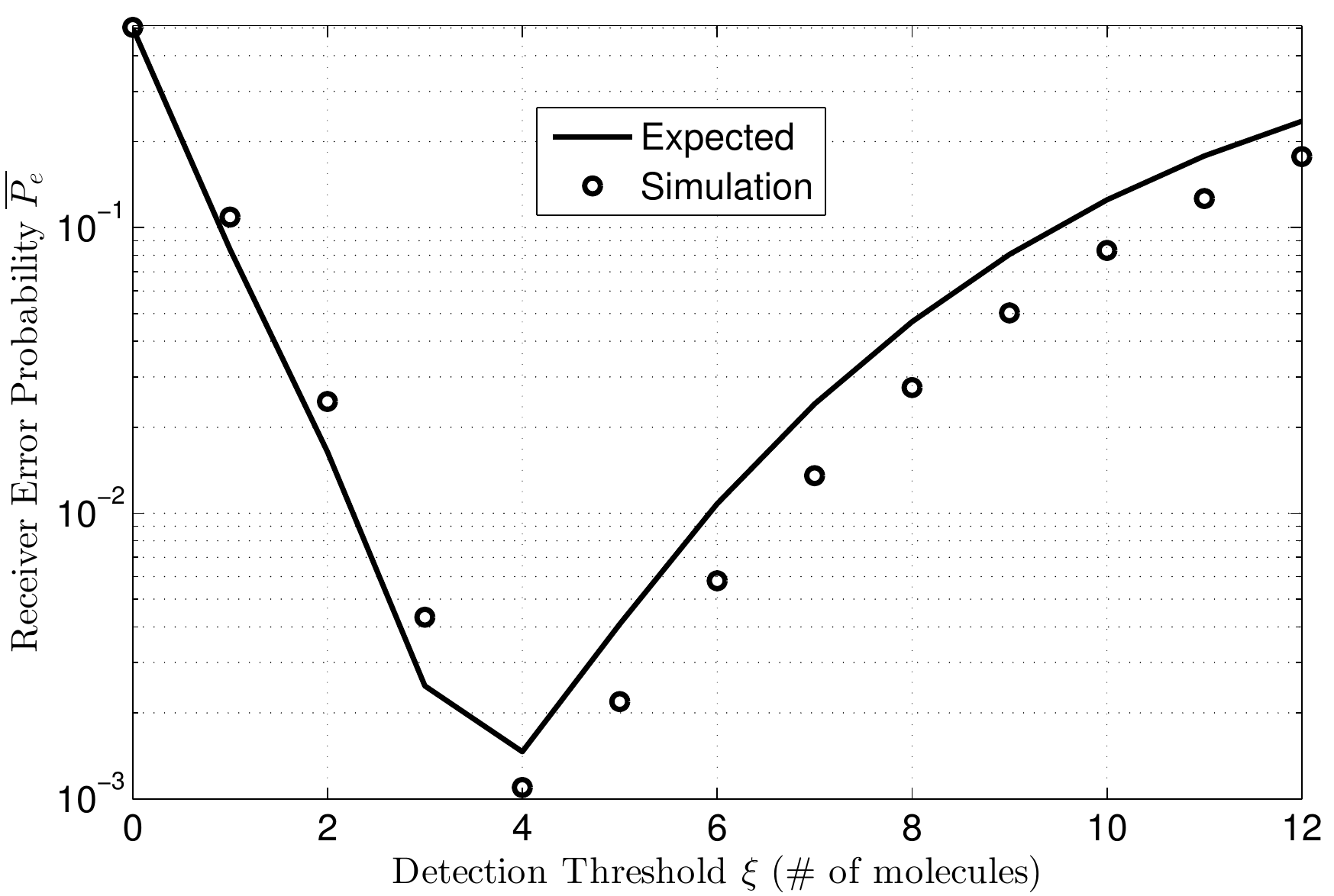}
	\caption{Evaluating the error probability of System 3 as a function of
	the bit decision threshold $\thresh$ at the receiver for bit interval
	$\T=120\,\mu\second$.}
	\label{fig_error_vs_thresh_system2}
	\end{figure}
}

\maketitle

\begin{abstract}
This paper studies the mitigation of intersymbol interference
in a diffusive molecular communication system using enzymes that freely diffuse
in the propagation environment.
The enzymes form reaction intermediates
with information molecules and then degrade them so that they cannot interfere
with future transmissions.
A lower bound expression on the expected number of
molecules measured at the receiver is derived.
A simple binary receiver detection scheme is proposed
where the number of observed molecules is sampled at the time
when the maximum number of molecules is expected. Insight is
also provided into the selection of an appropriate bit interval.
The expected bit error probability is derived as a function
of the current and all previously transmitted bits. Simulation
results show the accuracy of the bit error probability expression and the
improvement in communication performance by having active enzymes present.
\end{abstract}


\section{Introduction}

Recent interest in the design of nanonetworks, where communicating devices have
functional components that are on the order of nanometers in size, has emerged
for applications in areas such as biomedicine, environmental monitoring, and
manufacturing; see \cite{RefWorks:540, RefWorks:608}. The devices themselves
could share information over potentially longer distances, on the micrometer
scale and further. This communication capability is essential if the entire
devices are very small since they would have limited individual
processing capacity. Molecular communication is a nanonetwork design strategy
where a transmitter emits information molecules that are carried to an intended
receiver. It is a bio-inspired approach that can take
advantage of the many
mechanisms in cells and subcellular structures that already use the emission of
molecules for communication. By utilizing biological components, such as
genetically modified cells, we might hope to design networks that are
inherently biocompatible for implementation inside of living organisms.

The simplest propagation method in molecular communication
is free diffusion, which can be
modeled as a random walk. Molecules
that are released by a transmitter can
freely diffuse away without any external energy or infrastructure requirements.
Diffusion can be very fast over short distances, and is a common means of
communication in nature; many cellular processes rely on diffusion for limited
quantities of molecules to efficiently propagate both within and between
cells, as described in \cite[Ch. 16]{RefWorks:588}. Many researchers have also
adopted diffusion for the design of molecular communication networks, cf. e.g.
\cite{RefWorks:507, RefWorks:548, RefWorks:607, RefWorks:615, RefWorks:625,
RefWorks:525, RefWorks:534, RefWorks:574, RefWorks:609,
RefWorks:512, RefWorks:513, RefWorks:469,
RefWorks:671,RefWorks:643,RefWorks:677, RefWorks:667,
RefWorks:668,RefWorks:644,RefWorks:687}.

The average distance
travelled by a diffusing molecule is proportional to the square root of
the time that it takes to diffuse. So, molecular communication systems have to
deal with increasingly longer propagation times as the receiver is placed
further away.
The general lack of control over where molecules diffuse means that a large
number of molecules is required to ensure that a sufficient number arrive at
the receiver instead of diffusing away. Furthermore,
the receiver's ability to differentiate between the arrival of the same type
of molecule emitted at different times is reduced by how long it takes for
those molecules to leave the proximity of the transmitter and receiver. Unless
there is a mechanism in place to remove excess information molecules from the
environment, the transmission rate between a single transmitter and receiver is
limited by the on-going proximity of previously emitted molecules, i.e.,
intersymbol interference (\ISI).

The current literature on diffusion-based molecular communication has
primarily dealt with \ISI\, via passive strategies where the transmitter must
wait sufficiently long for previously-emitted information molecules to diffuse
away before it can release more molecules, thereby limiting the maximum
transmission rate. For example, \ISI\, has often been ignored, as in
\cite{RefWorks:507, RefWorks:548, RefWorks:607, RefWorks:615, RefWorks:625, RefWorks:671},
or it has been assumed that interfering molecules are released no earlier than the
previous bit interval, as in \cite{RefWorks:525, RefWorks:534, RefWorks:574,
RefWorks:609,RefWorks:643,RefWorks:677}.
\ISI\, from all previous transmissions has been
considered in $1$-dimensional diffusion environments in \cite{RefWorks:667,
RefWorks:668} and in $2$-dimensional environments in
\cite{RefWorks:644}, where the Viterbi algorithm is applied at the receiver
to optimally detect emissions from a transmitter that uses molecule shift
keying (though the Viterbi algorithm may be too complex for practical
implementation in small bio-inspired devices). An
upper-bound on the capacity of a 3-dimensional diffusive environment,
accounting for all \ISI, was recently derived in \cite{RefWorks:687}.

Communications
capacity can be significantly improved by adding a mechanism that actively
transforms information molecules so that they are no longer recognized by the
receiver. In general, chemical reactant molecules could perform this role, but
then they must be provided in stoichiometric excess relative to the information
molecules, otherwise their capacity to transform the information molecules will
degrade over time.
Catalysts, on the other hand, lower the activation energy for
biochemical reactions but do not appear in the stoichiometric expression of the
complete transformative reaction; unlike reactants, catalysts are not consumed.
Specifically, enzymes are catalytic biomolecules that can have the advantage of
very high selectivity for their substrates; see \cite[Ch. 16]{RefWorks:588}.
Some enzymes are already used in nature for the purpose of reducing \ISI; for
example, acetylcholinesterase is an enzyme in the neuromuscular junction that
hydrolyzes acetylcholine as it diffuses to its destination, as described in
detail in \cite[Ch. 12]{RefWorks:587}. Furthermore, another mechanism
regenerates acetylcholine at the transmitter from its hydrolyzed components so
that it can be re-used in future emissions.

We are interested in using enzymes
in the propagation environment due to their selectivity and because a single
enzyme can be recycled to react many times.
No additional complexity at either
the transmitter or receiver is required. The reduction in \ISI\, would
enable transmitters to release molecules more often, simultaneously increasing
the data rate and decreasing the probability of erroneous transmission. There
would also be less interference from neighbouring communication links, so
independent transmitter-receiver pairs could be placed closer together than in an
environment dominated by diffusion alone.

\ifOneCol
\else
	\figEnz{!tb}
\fi

The potential for information molecules to
participate in a chemical reaction mechanism has usually been considered only
at the receiver, as in \cite{RefWorks:512, RefWorks:625,RefWorks:723}.
Papers that have
considered information molecules reacting in the propagation environment include
\cite{RefWorks:513, RefWorks:469, RefWorks:711}.
In \cite{RefWorks:513}, the spontaneous
destruction and duplication of information molecules are treated as noise
sources but were not deliberately imposed to improve communication. In
\cite{RefWorks:469}, the exponential decay of information molecules was
considered via simulation as a method to reduce \ISI. However, information was
measured as the total number of molecules to reach the receiver, so the
achievable information rate actually decreased when information molecules were
allowed to decay. The placement of enzymes along the boundaries of the
propagation environment with the goal of reducing \ISI\,
was proposed in \cite{RefWorks:711} but analytical results were not provided.

In this paper, we present a model for analyzing diffusion-based molecular
communication systems when there are enzymes present throughout the entire
propagation environment. The enzymes
react with the information molecules via Michaelis-Menten kinetics,
which is a common mechanism for enzymatic reactions; see
\cite[Ch. 10]{RefWorks:585}. We first introduced this scenario in
\cite{RefWorks:631}, where we showed that enzymes
reduce the ``tail'' created when we rely on diffusion alone.
In Fig.~\ref{fig_enz}, we present a visual example of the
degradation of the diffusion ``tail'' when enzymes are present.
It is clear that, although enzymes reduce the expected
peak concentration, significantly less \ISI\, can be expected due to the
on-going degradation of information molecules throughout the propagation
environment. This paper expands
the work presented in \cite{RefWorks:631} and makes the following
contributions:
\begin{enumerate}
    \item As in \cite{RefWorks:631}, we present a lower bound expression
    on the expected number of information molecules observed at
    a receiver due to a
	transmitter that uses binary modulation to emit
	impulses of molecules when enzymes are present
	in the propagation environment.
    \item We derive the time at which the maximum number of information
    molecules is expected, both with and without enzymes.
    We also derive an upper bound expression on the time for the
    expected number of molecules to decrease from the maximum number to an
    arbitrary fraction of the maximum.
    This analytically shows that, for a given level of \ISI, a shorter bit
    interval can be achieved by adding enzymes, and provides insight
    into the selection of an appropriate bit interval.
    \item We design a simple detector where the receiver counts the
    number of molecules observed at the instant when the number of molecules
    is expected to be maximum and the observed number of molecules
    is compared to a binary decision threshold.
    \item We derive the bit error rate of this scheme for the
    first emission by the transmitter and then as a function of the current and
    all previous emissions.
	This derivation is the first to consider
    all \ISI\, in evaluating the bit error probability of a receiver
    in a 3-dimensional diffusive environment.
    \item As in \cite{RefWorks:631}, we justify a particle-based simulation
    framework and describe its implementation in our reaction-diffusion system.
\end{enumerate}

The rest of this paper is organized as follows. In Section~\ref{sec_model},
we introduce our system model for a single transmission link using
binary modulation,
including the degradation of information molecules via Michaelis-Menten
kinetics.
A lower bound expression on
the expected number of observed molecules at
the receiver when enzymes are present is derived in
Section~\ref{sec_perf} and compared with the baseline, no-enzyme scenario.
The performance analysis of the receiver, where we
calculate the signal degradation time and
derive the bit error rate for the simple detection scheme, is presented in
Section~\ref{sec_receiver}. In Section~\ref{sec_sim}, we describe the
simulation framework and provide some insight into the selection of appropriate
parameter values. In Section~\ref{sec_results}, we present and
discuss numerical and
simulation results. Conclusions and the on-going direction of
our research are described in Section~\ref{sec_concl}.

\section{Physical Model}
\label{sec_model}

There is a transmitter fixed at the origin of an unbounded
3-dimensional fluidic environment.
The receiver is a fixed spherical observer with radius
$\radmag{ob}$ and volume $\Vobs$. It is
centered at location $\{\x_0,\y_0,\z_0\}$ where
$\rad{0}$ is the vector from the origin to $\{\x_0,\y_0,\z_0\}$.
The receiver is a passive observer;
molecules can diffuse through it as they do through the entire environment.

There are three diffusive molecular species in the system that we are
interested in: $\A$ molecules, $\En$ molecules, and $\EA$ molecules.
$\A$ molecules are the information molecules that are released by the transmitter.
These molecules have a natural degradation rate that is negligible over the
time scale of interest, but they are able to act as substrates with enzyme
$\En$ molecules.
We apply Michaelis-Menten kinetics, which
is generally accepted as the
fundamental mechanism for enzymatic reactions (see \cite{RefWorks:587,
RefWorks:585}), to the $\A$ and $\En$
molecules:
\begin{equation}
\label{rxn_mechanism}
\En + \A \xrightleftharpoons[\kth{-1}]{\kth{1}} \EA
\xrightarrow{\kth{2}} \En + \AP,
\end{equation}
where $\EA$ is the intermediate formed by the binding of an $\A$ molecule to an
enzyme molecule and $\AP$ is the degraded (product) $\A$ molecule.
The reaction rate constants for the reactions as shown are
$\kth{1}$ in $\molecule^{-1}\metre^3\,\second^{-1}$,
$\kth{-1}$ in $\second^{-1}$, and $\kth{2}$ in $\second^{-1}$.
We see that $\A$ molecules are irreversibly
degraded by the reaction defined by $\kth{2}$
while the enzymes are released intact
so that they can participate in future reactions.
$\AP$ molecules are ignored once they are
formed because they cannot participate in future reactions and they are not
recognized by the receiver (we do not consider the re-generation of $\A$
molecules from $\AP$ molecules at the transmitter). Throughout
this paper, we refer to the three reactions in (\ref{rxn_mechanism})
associated with $\kth{1}$, $\kth{-1}$, and $\kth{2}$
as the binding, unbinding, and degradation reactions, respectively.

We use a common notation to refer to parameters of each molecular
species. We define these parameters for arbitrary species $\X$,
which could be either $\A$, $\En$, or $\EA$ molecules.
The number of molecules of species $\X$ is given by $\Nx{\X}$, and its
concentration at the point defined by vector $\rad{}$ and at time $t$ in
$\molecule\cdot\metre^{-3}$ is $\CxFun{\X}{\rad{}}{t}$.
For compactness, we will generally write $\CxFun{\X}{\rad{}}{t} = \Cx{\X}$.
We assume that every molecule of each species $\X$ diffuses independently of all
other molecules. We assume that all
free molecules are spherical in shape so that we can state that
each molecule diffuses
with diffusion constant $\Dx{\X}$, found using the Einstein relation as
\cite[Eq. 4.16]{RefWorks:587}
\begin{equation}
\label{JUN12_60}
\Dx{\X} = \frac{\bolt\temp}{6\pi \visc \Ri{\X}},
\end{equation}
where $\bolt$ is the Boltzmann constant ($\bolt = 1.38 \times 10^{-23}$ J/K),
$\temp$ is the temperature in kelvin, $\visc$ is the viscosity of the
medium in which the molecules are diffusing ($\visc \approx
10^{-3}\,\textnormal{kg}\cdot\metre^{-1}\second^{-1}$ for water at
$25\,^{\circ}\mathrm{C}$), and $\Ri{\X}$ is the molecule radius. Thus, the units
for $\Dx{\X}$ are $\metre^2/\second$. The diffusion of a single molecule along
one dimension has variance $2\Dx{\X} t$, where $t$ is the diffusing time
in seconds \cite[Eq. 4.6]{RefWorks:587}.

Communication occurs as follows. The transmitter emits impulses of
$\A$ molecules, where the number of molecules emitted is $\Nemit$.
This is a common emission
scheme in the molecular communication literature; see, for example,
\cite{RefWorks:443, RefWorks:512, RefWorks:614, RefWorks:615}. We deploy binary
modulation with constant bit interval $\T$, where $\Nemit$ molecules are
released at the start of the interval for binary 1 and no molecules are
released for binary 0. This method is also known as ON/OFF keying. $\Ne$
enzyme $\En$
molecules are randomly (uniformly) distributed throughout a finite cubic volume
$\Ve$ that includes both the transmitter (TX) and receiver (RX), as shown in
Fig.~\ref{venz} with the transmitter at the center.
The diffusion of the $\En$ molecules is restricted to within
$\Ve$, such that the \emph{total} concentration of the free and bound enzyme in
$\Ve$, $\Cx{\Etot}$, is constant and equal to $\Ne/\Ve$ (the local
concentration of free and bound enzyme in any subregion of $\Ve$ does vary over
time). Further details on the implementation of a finite $\Ve$ in the simulation
framework are provided in Section~\ref{sec_sim}, but we assume in our analysis
that $\Ve$ is infinite in size.

\ifOneCol
\else
	\figVenz{!tb}
\fi

The receiver counts the number of
free (unbound) $\A$ molecules that are within the receiver volume, without
disturbing those molecules. For a practical bio-inspired
system, the $\A$
molecules would need to bind to receptors on either the receiver surface or
within the receiver's volume, but we assume perfect passive counting in order
to focus on the propagation environment.

\section{Observations at the Receiver}
\label{sec_perf}

Generally, the spatiotemporal behavior of the information, enzyme, and
intermediate molecules can be
described using a system of reaction-diffusion partial differential equations.
Even though these equations are deterministic, they will enable stochastic
simulation. In this section, we use the reaction-diffusion
partial differential
equations to derive the expected number of information molecules at the receiver.

\subsection{Diffusion Only}

Fick's Second Law describes the motion of arbitrary
$\X$ molecules
undergoing independent diffusion as \cite[Ch.4]{RefWorks:587}
\begin{equation}
\label{APR12_21}
\pbypx{\Cx{\X}}{t} = \Dx{\X}\nabla^2\Cx{\X},
\end{equation}
for species $\X$ where $\Dx{\X}$ is the diffusion coefficient of the
species. Closed-form analytical
solutions for partial differential equations are not always possible and depend
on the boundary conditions that are imposed.
For comparison, we first consider the scenario of no
enzyme present, i.e., $\Ne = 0$. So, we immediately have
$\Cx{\En} = \Cx{\EA} = 0 \;\forall\, \rad{},t$, and
we only consider the diffusion of $\A$ molecules.
The \emph{expected} scaled (by $\Nemit$) impulse response at
a distance $\radmag{}$ from the transmitter is then \cite[Eq.
4.28]{RefWorks:587}
\begin{equation}
\label{APR12_22}
\Cx{\A} = \frac{\Nemit}{(4\pi \Da
t)^{3/2}}\EXP{\frac{-\radmag{}^2}{4\Da t}},
\end{equation}
where $t$ is the time since the $\Nemit$ information molecules
were released.
Eq. (\ref{APR12_22}) is the form that is typically used in molecular
communication to describe the local concentration at the receiver (where
$\rad{} = \rad{0}$, the vector from the transmitter to the center
of the receiver); the receiver is assumed to be a point observer, as in
\cite{RefWorks:498, RefWorks:546}, or
the concentration throughout the receiver volume is assumed to be uniform and
equal to that expected in the center, as in \cite{RefWorks:512}. Eq.
(\ref{APR12_22}) is the baseline against which we evaluate our proposed
system design.

\subsection{Reaction-Diffusion}

We now include active enzymes in our analysis.
The general reaction-diffusion equation for species $\X$ is
\cite[Eq. 8.12.1]{RefWorks:602}
\begin{equation}
\label{JUN12_33}
\pbypx{\Cx{\X}}{t} = \Dx{\X}\nabla^2\Cx{\X} + f\left(\Cx{\X},\rad{},t\right),
\end{equation}
where $f\left(\cdot\right)$ is the reaction term. Applying the principles
of chemical kinetics (see \cite[Ch. 9]{RefWorks:585}) to Michaelis-Menten
kinetics in (\ref{rxn_mechanism}), we write the complete
set of reaction-diffusion
equations for the species in our environment as
\begin{align}
\label{AUG12_39}
\pbypx{\Cx{\A}}{t} = &\; \Da\nabla^2\Cx{\A} -\kth{1}\Cx{\A}\Cx{\En} +
\kth{-1}\Cx{\EA}, \\
\label{AUG12_40}
\pbypx{\Cx{\En}}{t} = &\; \De\nabla^2\Cx{\En} -\kth{1}\Cx{\A}\Cx{\En} +
\kth{-1}\Cx{\EA} + \kth{2}\Cx{\EA},\\
\label{AUG12_41}
\pbypx{\Cx{\EA}}{t} = &\; \Da\nabla^2\Cx{\EA} +\kth{1}\Cx{\A}\Cx{\En} -
\kth{-1}\Cx{\EA} - \kth{2}\Cx{\EA}.
\end{align}

This system of equations is highly coupled due to the reaction terms and has no
closed-form analytical solution under our boundary conditions; we must make some
simplifying assumptions:
\begin{enumerate}
    \item We assume that the degradation reaction
	is relatively very fast, i.e., $\kth{2} \to\infty$.
	\item We assume
	that the unbinding reaction is relatively very slow, i.e.,
	$\kth{-1} \to 0$.
\end{enumerate}

From the first assumption, we can claim that $\Cx{\En}$ remains
close to the total concentration of free and bound enzyme,
i.e., $\Cx{\Etot} = \Ne/\Ve$,
over all time and space, since there will never be a significant quantity
of bound enzyme. Thus, $\Cx{\EA}$ remains small
over all time and space. Before applying explicit bounds on $\Cx{\En}$
and $\Cx{\EA}$, it is sufficient for a solution to assume that they are both steady
and uniform (i.e., they are constant) and it can then be shown that,
in our system, (\ref{AUG12_39}) has solution
\begin{equation}
\label{JUN12_47_proof}
\Cx{\A} \approx \frac{\Nemit}{(4\pi \Da
t)^{3/2}}\EXP{-\kth{1}\Cx{\En}t - \frac{\radmag{}^2}{4\Da t}} +
\kth{-1}\Cx{\EA}t,
\end{equation}
and we ignore (\ref{AUG12_40}) and (\ref{AUG12_41}). Next, we
apply the upper bound on $\Cx{\En}$ (i.e., $\Cx{\Etot}$) and
use the second assumption to apply a lower bound on
$\kth{-1}\Cx{\EA}$ (i.e., 0) to write the lower bound on the expected impulse
response as
\begin{equation}
\label{JUN12_47}
\Cx{\A} \ge \frac{\Nemit}{(4\pi \Da
t)^{3/2}}\EXP{-\kth{1}\Cx{\Etot}t - \frac{\radmag{}^2}{4\Da t}},
\end{equation}
which is intuitively a lower bound because the actual degradation due to enzymes
can be no more than if all enzymes were always unbound. The tightness
of this lower bound depends directly on the accuracy of our two assumptions
about the reaction rates. The actual expected concentration will be between
(\ref{JUN12_47}) and the diffusion-only case (\ref{APR12_22}), but will
become closer to (\ref{JUN12_47}) if the two assumptions are more
accurate (i.e., if the assumptions are not accurate, then the mitigation
of \ISI\, by adding enzymes is less than expected).
In general, this lower bound
loses accuracy as $\EA$ is initially created
($\Cx{\En} < \Cx{\Etot}$),
but it eventually improves with time for non-zero reaction rates
as all $\A$ molecules
are degraded and none remain to bind with the enzymes
($\Cx{\A}, \Cx{\EA} \to 0$,
$\Cx{\En} \to \Cx{\Etot}$, as $t \to \infty$).

An alternate solution for (\ref{AUG12_39}) can be derived without our
two assumptions about the reaction rate constants $\kth{-1}$ and $\kth{2}$,
where it is only assumed that $\Cx{\EA}$ is constant. This is a common
step in the analysis of Michaelis-Menton kinetics; see
\cite[Ch. 10]{RefWorks:585} and its use when
considering enzymes only at the receiver in \cite{RefWorks:625}. The
resulting expression is similar to (\ref{JUN12_47}), where the binding rate
$\kth{1}$ is replaced with $\kth{1}\kth{2}/\left(\kth{-1} + \kth{2}\right)$,
but it is an \emph{approximation} and not a lower bound.

Eq. (\ref{JUN12_47}) can be
directly compared with (\ref{APR12_22}). The presence of enzyme
molecules results in an
additional decaying exponential term. This decaying exponential is what will
eliminate the ``tail'' that is observed under diffusion alone. An immediate
result from (\ref{JUN12_47}) is that increasing either the binding rate
$\kth{1}$ or the total enzyme concentration $\Cx{\Etot}$
will result in a faster-decaying ``tail'' and thus decrease \ISI, albeit
at the cost of also decreasing the useful signal in the desired bit interval.

We will assume throughout the remainder of this paper that the expected
concentration of information $\A$ molecules within the receiver is
uniform and equal to that expected at the center of the receiver.
We studied the accuracy of this assumption in \cite{RefWorks:706},
where we showed it is accurate for a receiver that
is sufficiently far from the transmitter.
We have already established that the receiver is able to count the number of
free $\A$ molecules that are within the receiver volume, so we can readily
convert the expected concentration into the expected number of
observed $\A$ molecules at the receiver, $\Nobsavgt$.
Using (\ref{JUN12_47}) with enzymes and (\ref{APR12_22})
without enzymes we can write
\begin{equation}
\label{AUG12_59}
\Nobsavgt = \CxFun{\A}{\rad{0}}{t}\Vobs,
\end{equation}
for the expected number of observed molecules,
where $\rad{0}$ is the vector from the origin to the center of the receiver.

\section{Receiver Performance Analysis}
\label{sec_receiver}

In this section, we first consider the signal degradation with enzymes and
provide a method to calculate an appropriate bit interval $\T$. Then, we derive
the bit error rate of the considered receiver
and provide approximations that facilitate
closed-form expressions. We design the reception mechanism such
that the receiver counts the number of free $\A$ molecules observed within
the receiver volume
$\Vobs$ at the instant when the expected number of molecules is maximal
(assuming that the transmitter emits molecules at the start of the bit
interval). A single decision threshold is used for the receiver to determine
whether a binary 1 or binary 0 was sent by the transmitter.
This is a relatively simple reception mechanism that facilitates
analysis and approximates a physically realizable scheme. For example,
if a biochemical response mechanism were triggered at the receiver
when the information molecule concentration reached a threshold level,
then the threshold is most likely to be exceeded when the maximum number
of information molecules is expected. The more realistic case, where the
threshold could be exceeded at any time, is an interesting problem
that we leave for future work.

\subsection{Signal Degradation with Enzymes}
From (\ref{JUN12_47}) and (\ref{AUG12_59}),
the number of information $\A$ molecules expected at the receiver is
\begin{equation}
\label{JUN12_47_mult}
\Nobsavg{t} \ge \frac{\Vobs\Nemit}{(4\pi \Da
t)^{3/2}}\EXP{-\kth{1}\Cx{\Etot}t - \frac{\radmag{0}^2}{4\Da t}},
\end{equation}
assuming that the transmitter releases the $\Nemit$ information $\A$
molecules at $t =
0$. It is straightforward to take the derivative of (\ref{JUN12_47_mult}) with
respect to $t$ to find the time, $\tmax$, at which the
maximum number of molecules is expected, found as
\begin{equation}
\label{JUN12_68}
\tmax = \frac{-3 + \sqrt{9 +
\left(4\kth{1}\Cx{\Etot}\radmag{0}^2\right)
/\Dx{\A}}}{4\kth{1}\Cx{\Etot}},
\end{equation}
where we only consider non-negative finite time. The maximum number of expected
molecules at the receiver, $\Nobsavgmax$, is then found by
substituting (\ref{JUN12_68}) into (\ref{JUN12_47_mult}). By comparison,
when there are no enzymes present, i.e., $\Cx{\Etot} = 0$,
we substitute (\ref{APR12_22}) instead of (\ref{JUN12_47}) into
(\ref{AUG12_59}) and find that
the maximum number
of molecules is expected at time
\begin{equation}
\label{JUN12_73}
\tmax\Big|_{\Cx{\Etot} = 0} = \frac{\radmag{0}^2}{6\Dx{\A}},
\end{equation}
and it is straightforward to show that, for all valid (i.e., non-negative)
parameter values, $\tmax \le \tmax\big|_{\Cx{\Etot} =
0}$. So, when enzymes are added, the maximum number of molecules is expected no
later than when enzymes are not added.

By inspection of (\ref{JUN12_47_mult}), we
also observe that, for a given $t$, we expect to observe
fewer $\A$ molecules when enzymes are present, i.e., $\Nobsavg{t} \le
\Nobsavg{t}\big|_{\Cx{\Etot} = 0}, \forall\,t$. Thus, we immediately
have that $\Nobsavgmax \le \Nobsavgmax\big|_{\Cx{\Etot} = 0}$. In addition,
when enzymes are present, the expected number of molecules will decrease
to any value sooner than when enzymes are absent, i.e., \ISI\, must decrease.
In order to consider the
selection of an appropriate bit interval $\T$, we are interested in solving for
the time required for $\Nobsavg{t}$ to decrease to some threshold value,
i.e., find $\threshT$ that satisfies
\begin{equation}
\label{JUN12_75}
\frac{\Nobsavg{\threshT}}{\Nobsavgmax} \le \threshInterval,
\end{equation}
where $0 < \threshInterval < 1$ is a threshold fraction of the maximum expected
number of molecules and $\threshT > \tmax$.
Solving (\ref{JUN12_75}) presents two challenges.
First, it cannot be strictly satisfied because we only have a
lower bound on $\Nobsavg{t}$ when enzymes are present;
showing that the lower bound is lower than
$\threshInterval\Nobsavgmax$ does not satisfy (\ref{JUN12_75}).
However, we will assume that the system parameters are such that
the lower bound (\ref{JUN12_47_mult}) is met
with equality; i.e., we assume that $\kth{2} \to \infty$ and $\kth{-1} \to 0$.
Second, an
analytical solution to (\ref{JUN12_75}) using
the lower bound is not possible; further bounding
will be required to obtain a closed-form expression for $\threshT$.
Alternatively,
we can solve (\ref{JUN12_75}) numerically using (\ref{JUN12_47_mult})  without
any further bounding by initializing $\threshT = \tmax$ and gradually
increasing $\threshT$ by increments much smaller than $\tmax$ until
the inequality in (\ref{JUN12_75}) is satisfied. The numerical solution
enables us to assess the accuracy of a closed-form expression.

We derive a bound for $\threshT$ that satisfies (\ref{JUN12_75})
by first observing that, for smaller values of $\threshT$ (and over the
range that we are interested in), the decay
of (\ref{JUN12_47_mult}) is dominated by ${\threshT}^{-\frac{3}{2}}$
and not the exponential. Therefore, we propose applying an upper bound
on the exponential term by replacing it with its maximum value.
It can be shown that the exponential term in (\ref{JUN12_47_mult})
has maximum value $\EXP{\radmag{0}\sqrt{\kth{1}\Cx{\Etot}/\Dx{\A}}}$ when $\threshT
= \radmag{0}/\sqrt{4\kth{1}\Cx{\Etot}\Dx{\A}}$. We replace the exponential
term with this maximum, substitute (\ref{JUN12_47_mult}) into
(\ref{JUN12_75}), and solve for $\threshT$ as
\begin{equation}
\label{JUN12_77}
\threshT \ge
\frac{1}{4\pi\Dx{\A}}
\left(\frac{\Vobs\Nemit}{\threshInterval\Nobsavgmax}\right)^\frac{2}{3}
\EXP{-\frac{2}{3}\radmag{0}\sqrt{\frac{\kth{1}\Cx{\Etot}}{\Dx{\A}}}}.
\end{equation}

In addition, we can guarantee that
$\threshT > \tmax$ because we used the upper-bound on the
exponential term. Similarly, for the case without enzymes present, we
replace the exponential term with its maximum value of $1$ and
find that
\begin{equation}
\label{JUN12_80}
\threshT\Big|_{\Cx{\Etot} = 0} \ge
\frac{1}{4\pi\Dx{\A}}
\left(\frac{\Vobs\Nemit}{\threshInterval\Nobsavgmax\big|_{\Cx{\Etot} =
0}}\right)^\frac{2}{3}
\end{equation}
satisfies (\ref{JUN12_75}). A strict comparison between
(\ref{JUN12_77}) and (\ref{JUN12_80}) is not fair due to the
challenges previously mentioned. However,
these expressions do provide some guidance
in the selection of an appropriate bit interval.
We will generally assume that the bit interval time $\T$ is sufficiently
long so that $\T > \tmax$.

\subsection{Error Rate at the Receiver}

In our simple detection scheme, the receiver counts the number of free
information $\A$
molecules at time $\tmax$ after the start of the bit interval and compares that
number with decision threshold $\thresh$. We assume that there is perfect
synchronization between the transmitter and receiver to emphasize the
limitations of intersymbol interference.
As noted at the beginning of this section, our proposed detector
approximates a physically realizable detector; we do not expect to easily
achieve perfect synchronization between devices.
Recalling that $\T$ is the bit
interval time, the decision sampling time for the $j$th bit interval is
$(j-1)\,\T+\tmax$.

Let $\data{j}$ be the $j$th information bit sent by the
transmitter, i.e., sent at the beginning of the $j$th bit interval, and let
the \emph{a priori} probabilities of the transmitted bits be
$\Pr(\data{j} = 1) = \Pone$ and $\Pr(\data{j} = 0) = \Pzero = 1-\Pone$, where
$\Pr(\cdot)$ denotes probability.
Let $\dataObs{j}$ be the $j$th received bit at
the receiver. Thus, the reception mechanism can be written as
\begin{equation}
\label{SEP12_04}
\dataObs{j} = \left\{
 \begin{array}{rl}
  1 & \text{if } \Nobst{(j-1)\,\T+\tmax} \ge \thresh,\\
   0 & \text{if } \Nobst{(j-1)\,\T+\tmax} < \thresh.
 \end{array} \right.
\end{equation}

It is clear that an error occurs if $\data{j} \neq \dataObs{j}$, and we define
the error probability of the $j$th bit $\Pe{j} = \Pr(\data{j} \neq
\dataObs{j})$. So, we are interested in evaluating
$\Pr\left(\Nobst{(j-1)\,\T+\tmax} \ge \thresh\right)$, a
function of the current and all previous emissions by the transmitter. We begin
by considering the first bit, i.e., $j = 1$, and then extend the result to any
$j$th bit in the transmission. Generally, bits transmitted later will have
a higher probability of being detected in error because there are more previous
bits to create \ISI.

Consider the first bit for
the case $\data{1} = 1$ (for the case $\data{1} = 0$ there are
no information
$\A$ molecules anywhere in the system at the time $\tmax < \T$
and so there will be none observed at
the receiver). The lower bound on the expected number of observed molecules is
also a lower bound on the probability density function (\PDF) over all time and
space for a single molecule if we set $\Nemit = 1$ and assume that the
location and state of any one $\A$ molecule is independent of the other $\A$
molecules. Therefore, a lower bound on the probability $\Pobsx{t}$ that a given
molecule is observed within the receiver volume $\Vobs$ at
time $t$ is found by integrating
(\ref{JUN12_47}) over $\Vobs$.
However, we recall from (\ref{AUG12_59}) that we simplified the integration
by assuming that the concentration
of molecules within the receiver is uniform and equal to that expected
at the center of the receiver.
Thus, we write
\begin{equation}
\label{AUG12_59_not_DMLS}
\Pobsx{t} \ge
\frac{\Vobs}{(4\pi \Da t)^{3/2}}\EXP{-\kth{1}\Cx{\Etot}t -
\frac{\radmag{0}^2}{4\Da t}},
\end{equation}
and we will assume that (\ref{AUG12_59_not_DMLS}) is met with equality,
i.e., we assume that $\kth{2} \to \infty$ and $\kth{-1} \to 0$.
Generally, we have $\Nemit$ information molecules,
and each molecule is either inside
$\Vobs$ at a given time or outside, so the number of observed molecules follows
the binomial distribution. Thus, we can write \cite[Ch. 3]{RefWorks:635}
\begin{multline}
\label{JUN12_81}
\Pr(\Nobst{t} \ge \thresh) = \\
\sum_{\w=\thresh}^{\Nemit}\!
\binom{\Nemit}{\w}\Pobsx{t}^\w \!\left(1-\!\Pobsx{t}\right)^{\Nemit-\w}.
\end{multline}

Eq. (\ref{JUN12_81}) is exact for a given $\Pobsx{t}$ but is difficult to
evaluate for large values of $\Nemit$. However, as noted in
\cite{RefWorks:525}, we can write (\ref{JUN12_81}) in an equivalent form as
\begin{equation}
\label{JUN12_82}
\Pr(\Nobst{t} \ge \thresh) = \Ix{\Pobs}{\thresh, \Nemit-\thresh+1},
\end{equation}
where $\Ix{\Pobs}{\cdot, \cdot}$ is the regularized incomplete beta function
based on individual probability $\Pobsx{t}$,
i.e., \cite[Eq. 8.392]{RefWorks:402}
\begin{equation}
\Ix{\Pobs}{a, b} = \frac{\int\limits_0^{\Pobs}t^{a-1}(1-t)^{b-1}dt}
{\int\limits_0^1 t^{a-1}(1-t)^{b-1}dt}.
\end{equation}

Furthermore, we also note that
\ifOneCol
\begin{IEEEeqnarray}{r}
\Pr(\Nobst{t} = \w) = \label{JUN12_87}
\Ix{\Pobs}{\w, \Nemit-\w+1} - \: \Ix{\Pobs}{\w+1, \Nemit-\w}.
\end{IEEEeqnarray}
\else
\begin{IEEEeqnarray}{rCl}
\Pr(\Nobst{t} = \w) & = & \label{JUN12_87}
\Ix{\Pobs}{\w, \Nemit-\w+1} \nonumber\\
&& - \: \Ix{\Pobs}{\w+1, \Nemit-\w}.
\end{IEEEeqnarray}
\fi

Eqs. (\ref{JUN12_82})
and (\ref{JUN12_87}) can be evaluated numerically but the incomplete beta
function does not easily lend itself to optimization. For example, its
derivative cannot be written in closed form. We consider two approximations of
the binomial distribution. For infinitely large $\Nemit$ and infinitely small
$\Pobsx{t}$, such that their product is a finite positive number, the
binomial distribution approaches the Poisson distribution with mean
$\Nobsavg{t} = \Nemit\Pobsx{t}$, and we can write \cite[Ch. 3]{RefWorks:635}
\begin{equation}
\label{JUN12_87_Poiss}
\Pr(\Nobst{t} \!=\! \w)\poissBar \!= 
\frac{{\Nobsavg{t}}^\w\EXP{-\Nobsavg{t}}}{\w!},
\end{equation}
and so
\begin{equation}
\label{JUN12_81_Poiss}
\Pr(\Nobst{t} \ge \thresh)\poissBar \!=\!
1-\EXP{-\Nobsavg{t}}
\!\sum_{\w = 0}^{\thresh-1}\!\frac{{\Nobsavg{t}}^\w}{\w!}.
\end{equation}

Alternatively, we can approximate the binomial distribution with a Gaussian
distribution with mean $\Nobsavg{t}$ and variance
$\Nobsavg{t}\left(1-\Pobsx{t}\right)$. This approximation has been applied
by other authors for molecular communication, cf. e.g. \cite{RefWorks:534,
RefWorks:512}, and is valid when $\Pobsx{t}$ is not close to one
or zero and $\Nobsavg{t}$ is sufficiently large. Generally, this
approximation will not be as accurate as using the Poisson distribution because
we will tend to have very small values for $\Pobsx{t}$. We still consider
the Gaussian distribution because it does not include any factorials and so can
be more computationally efficient than the Poisson distribution. The Gaussian
approximation enables us to write
\begin{equation}
\label{JUN12_89}
\Pr(\Nobst{t} = \w)\gaussBar = 
\frac{\EXP{\frac{\left(\w-\Nobsavg{t}\right)^2}
{2\Nobsavg{t}\left(1-\Pobsx{t}\right)}}}
{\sqrt{2\pi\,\Nobsavg{t}\left(1-\Pobsx{t}\right)}},
\end{equation}
and by using the error function \cite[p. 406]{RefWorks:414} we can show that
\ifOneCol
\begin{equation}
\label{JUN12_90}
\Pr(\Nobst{t} \ge \thresh)\gaussBar = \frac{1}{2}\left[1-
\ERF{\frac{\thresh-\Nobsavg{t}}
{\sqrt{2\,\Nobsavg{t}\left(1-\Pobsx{t}\right)}}}\right].
\end{equation}
\else
\begin{multline}
\label{JUN12_90}
\Pr(\Nobst{t} \ge \thresh)\gaussBar \\ = \frac{1}{2}\left[1-
\ERF{\frac{\thresh-\Nobsavg{t}}
{\sqrt{2\,\Nobsavg{t}\left(1-\Pobsx{t}\right)}}}\right].
\end{multline}
\fi

To evaluate the error probability for the first bit, we use either of
(\ref{JUN12_82}), (\ref{JUN12_81_Poiss}), or (\ref{JUN12_90}). As noted, an
error in $\dataObs{1}$ is possible only when $\data{1} = 1$. Thus, the
probability of error in the first bit is
\begin{equation}
\label{SEP12_03}
\Pe{1} = \Pone\left[1 - \Pr(\Nobst{\tmax} \ge \thresh)\right].
\end{equation}

The probability of error of the $j$th bit is
a function of all of the first $j$ bits, since information
$\A$ molecules can remain in the
propagation environment from any of the previous emissions by the transmitter.
The common practice when deriving error rates in the molecular
communication literature is to assume that information $\A$
molecules remain in the
environment for no more than two transmission intervals, cf. e.g.
\cite{RefWorks:525, RefWorks:534, RefWorks:574, RefWorks:609, RefWorks:625}.
We make no assumptions about how long information molecules
remain in the proximity of the receiver in order to present
a general bit error expression that includes \emph{all} \ISI, which has
not yet been developed in the literature for a 3-dimensional
environment.
We will assume that the number of $\A$ molecules, observed at some time
$t$, that were emitted at the start of the $j$th
bit interval, are \emph{independent} of the number
of molecules, also observed at that time $t$, that were emitted
at the start of any other bit interval. We note that this is not strictly true
because an $\En$ molecule that is bound to an $\A$ molecule is temporarily
unavailable to bind to the $\A$ molecules of other transmissions (the mean time
of unavailability is controlled by the value of degradation rate $\kth{2}$,
but we have assumed for
analysis that $\kth{2} \to \infty$).

From the independence of molecules emitted at the start of every
bit interval, the number of molecules observed at the receiver at time $t$
is simply the sum of the number of molecules observed due to every emission.
We emphasize that any molecule observed within the receiver volume
$\Vobs$ could have been emitted
during the current or any previous bit interval.
We define $\Nobsn{j} = \Nobst{(j-1)\T+\tmax}$, i.e., all information $\A$
molecules observed
at time $\tmax$ within the $j$th bit interval, where $\T$ is the bit interval
time, and write
\begin{equation}
\Nobsn{j} = \sum_{i=1}^j \Nobsn{j;i},
\end{equation}
where
$\Nobsn{j;i}$ is the number of molecules observed at the time $\tmax$ in
the $j$th bit interval
that were emitted at the start of the $i$th bit interval (we note
that $\Nobsn{j;i} = 0, \forall j$ if $\data{i} = 0$). Thus, $\Nobsn{j}$
is a random variable that is a sum of random variables.
From \cite[Ch. 5]{RefWorks:725}, a random variable that is a sum of
all Binomial, all Poisson, or all Gaussian random variables is also
a Binomial, Poisson, or Gaussian random variable, respectively, and
its mean is the sum of the means of the individual variables. We can
then immediately write the expected number of molecules observed
at the receiver at time $\tmax$ within the $j$th bit interval as
\begin{equation}
\Nobsavgn{j} = \Nemit\sum_{i=1}^j \data{i}\Pobsx{(j-i)\T+\tmax},
\label{EQ13_04_02}
\end{equation}
where $\Nemit$ is the number of molecules released
when $\data{i} = 1$, and $\Pobsx{t}$ from (\ref{AUG12_59_not_DMLS})
is the probability that a single molecule is observed within $\Vobs$
at time $t$ after its release from the transmitter.
Given the mean $\Nobsavgn{j}$, $\Pr\left(\Nobsn{j} = \w\right)$ can be
evaluated from (\ref{JUN12_87_Poiss}) using Poisson statistics or
from (\ref{JUN12_89}) using Gaussian statistics,
where $\Nemit\Pobsx{t}$ is replaced with $\Nobsavgn{j}$ (it is also
possible to use Binomial statistics, but the adaptation of (\ref{JUN12_87})
is less straightforward so we omit the details here).
Analogously,
$\Pr\left(\Nobsn{j} \ge
\thresh\right)$ can be evaluated from (\ref{JUN12_81_Poiss}) using Poisson
statistics
or (\ref{JUN12_90}) using Gaussian statistics.

Given $\Pr\left(\Nobsn{j} \ge \thresh\right)$, we can immediately
evaluate $\Pe{j}$ for a given transmitter bit sequence
$\data{i}, i \in \{1,2,\ldots,j\}$. Given the transmitter
sequence for all prior intervals, i.e., $i < j$,
the probability of error in the $j$th bit is
\begin{IEEEeqnarray}{rCl}
\IEEEeqnarraymulticol{3}{l}{
\Pe{j} = \Pone\Pr\left(\Nobsn{j} <
\thresh \,|\, \data{j} = 1, \data{i}, i < j\right)
} \nonumber\\ \quad\quad\;\; & + &
\Pzero\Pr\left(\Nobsn{j} \ge\thresh \,|\, \data{j} = 0, \data{i}, i < j\right).
\IEEEeqnarraynumspace
\label{SEP12_07}
\end{IEEEeqnarray}

If we have \emph{a priori} knowledge of the current and all previous bits,
then the expected probability of error is found by taking the appropriate
term in (\ref{SEP12_07}). Generally, assuming no \emph{a priori} knowledge,
we must evaluate (\ref{SEP12_07}) for all $2^{j-1}$ possible prior
bit sequences, though in practice we can average the expected error
probability over a subset of all possible sequences.

The common practices in the literature of either ignoring \ISI\,
or only considering the interference caused by emission in the
previous bit interval, cf. e.g.
\cite{RefWorks:525, RefWorks:534, RefWorks:609, RefWorks:625}, can both be
evaluated as special cases of (\ref{SEP12_07}) by limiting the number of
terms used in finding $\Nobsavgn{j}$ in (\ref{EQ13_04_02}).
Specifically, we initialize $i = j$ to
ignore \ISI\, and initialize $i = j-1$ to
only consider \ISI\, from the previous bit interval.

\section{Simulation Framework}
\label{sec_sim}

This section describes the framework used to perform stochastic
simulations of the system of reaction-diffusion equations described by
(\ref{AUG12_39})-(\ref{AUG12_41}), which also simplifies to the
diffusion-only case if the total enzyme concentration $\Cx{\Etot} = 0$.

\subsection{Choice of Framework}

Our simulation framework uses a particle-based method,
where the precise locations of all individual molecules are known.
The primary alternative, subvolume-based methods,
divide the environment into subvolumes and each molecule is
known to be in a given subvolume.
Particle-based methods tend to be less computationally efficient than
subvolume-based methods, but they do not have to meet the latter's well-stirred
requirement, where every subvolume should have many more nonreactive molecular
collisions than reactive collisions, as described in
\cite{RefWorks:616, RefWorks:612}.
A general criterion for subvolume size is that the typical
diffusion time for each species should be much less than the typical reaction
time; see \cite{RefWorks:613}. In order to satisfy this criterion, we would need
to use very small subvolume sizes relative to the total size of the environment
under consideration. If we used such small subvolumes, then we would not
gain in computational efficiency and, for small (nanoscale) environments, the
subvolume size would not be much greater than the size of individual molecules.
Thus, we proceed with a particle-based method.

Every
free molecule in a particle-based method
diffuses independently along each dimension. Such methods require
a constant global time step $\Delta t$ (the chosen value
of $\Delta t$ represents a tradeoff in accuracy
and simulation time) and there is a separation in the
simulation of reaction and diffusion; see \cite{RefWorks:623}.
First, all free
molecules are independently displaced
along each dimension by generating normal
random variables with variance $2\Dx{\X} \Delta t$, where $\Dx{\X}$
is the diffusion coefficient of arbitrary species $\X$.
Next, potential reactions are
evaluated to see whether they would have occurred during $\Delta t$. For
bimolecular reactions, a binding radius $\radbind$ is defined as how close
the centers of two reactant molecules need to be at the end of $\Delta t$ in
order to assume that the two molecules collided and bound during $\Delta t$.
For unimolecular reactions, a random number is generated using the
corresponding rate
constant to declare whether the reaction occurred during $\Delta t$.

\subsection{Simulating Reactions}

All three reactions in (\ref{rxn_mechanism}) have an enzyme $\En$
molecule as a reactant and an intermediate $\EA$ molecule as a
product (or vice versa). Thus,
we must jointly consider the two unimolecular reactions with $\EA$ as the reactant,
and we must take care when modeling the binding and unbinding
reactions so that the binding reaction does not occur when not intended.

The probability of the unbinding reaction ($\kth{-1}$) occuring
is a function of both the unbinding and degradation rate constants, written
as \cite[Eq. 14]{RefWorks:623}
\begin{equation}
\label{AUG12_23}
\Pr(\textnormal{Reaction $\kth{-1}$}) =
\frac{\kth{-1}}{\kth{-1}+\kth{2}}
\left[1 - \EXP{-\Delta t \left(\kth{-1}+\kth{2}\right)}\right],
\end{equation}
and the degradation reaction ($\kth{2}$) has
an analogous expression by swapping $\kth{-1}$ and $\kth{2}$. A single
random number uniformly distributed between 0 and 1 can then be used to
determine whether a given $\EA$ molecule reacts, and, if so, which reaction
occurs.

The bimolecular binding
reaction
(i.e., the binding of an enzyme $\En$ molecule
and an information
$\A$ molecule at rate $\kth{1}$ to form an intermediate $\EA$ molecule) is reversible,
so we must be careful in our choice of
binding radius $\radbind$, time step $\Delta t$, and what we assume
when $\EA$ reverts back to $\En$ and $\A$ molecules.
If the $\En$ and $\A$ molecules are not physically separated when
the unbinding reaction ($\kth{-1}$) occurs,
and $\radbind$ is much larger than the expected separation of the
$\En$ and $\A$ molecules by diffusion in the following time step,
then the binding reaction will very likely occur in the next time step
regardless of the actual value of $\radbind$.
Therefore, we must consider the root mean square of the separation
of $\En$ and $\A$ molecules in a single time step,
given as \cite[Eq. 23]{RefWorks:623}
\begin{equation}
\label{AUG12_25}
\stepl = \sqrt{2\left(\Da + \De\right)\Delta t},
\end{equation}
where $\Da$ and $\De$ are the constant diffusion coefficients
of the $\A$ and $\En$ molecules, respectively.
In general, an unbinding radius that is larger than $\radbind$
is defined to separate the two molecules as soon as the reversible
unbinding reaction occurs. The objective in doing so is
to prevent the automatic re-binding of the same two molecules
in the next time step and
more accurately model the reaction kinetics; see \cite{RefWorks:623}.
However, if $\stepl \gg \radbind$, i.e., if the expected separation
of the two molecules in one time step is much larger than the binding radius, then
an unbinding radius is unnecessary and it is sufficient to keep
the $\A$ and $\En$ molecules at the same coordinates when the unbinding reaction
occurs. If $\Delta t$ is sufficiently large,
then we can define $\radbind$ as \cite[Eq.
27]{RefWorks:623}
\begin{equation}
\label{AUG12_26}
\radbind = \left(\frac{3\kth{1}\Delta t}{4\pi}\right)^\frac{1}{3},
\end{equation}
and this expression is only valid if $\stepl \gg \radbind$. Thus,
if we are careful with our selection of $\kth{1}$ and $\Delta t$,
such that $\stepl \gg \radbind$ with $\radbind$ given by (\ref{AUG12_26}),
then we can legitimately use (\ref{AUG12_26}) to define $\radbind$.
If $\stepl \gg \radbind$ is not satisfied,
then $\radbind$ must be found using numerical
methods; see \cite{RefWorks:623}. In our simulations,
we ensure that the use of (\ref{AUG12_26}) is justified.

When a pair of
$\A$ and $\En$ molecules are within $\radbind$ of each other,
we move both of them to the midpoint of the line
between their centers and re-label them as a single
$\EA$ molecule. If the corresponding unbinding reaction occurs
in a later time step, then the molecule is re-labeled as separate $\A$ and
$\En$ molecules and we do not change their locations until they
diffuse in the following time step.

\subsection{Simulating the Transmitter and Receiver}

We simulate emissions at the transmitter by initializing $\Nemit$ $\A$
molecules centered at the origin and with a separation of $2\Ri{\A}$ between
adjacent molecules so that together they form a spherical shape. The receiver
can make observations only at integer multiples of the time step
$\Delta t$, so for detection we round $\tmax$ to the nearest multiple of
$\Delta t$. When an observation is made, all free $\A$ molecules whose centers
are within $\Vobs$ are counted.

\subsection{Simulating an Unbounded Environment}

We noted in Section~\ref{sec_model} that the physical environment
is unbounded but the movement of enzyme $\En$ molecules is restricted to
the large volume
$\Ve$. We force the enzymes to stay within $\Ve$ by reflecting them
off of the boundary of $\Ve$ if diffusion carries them outside. In doing so,
we simulate a uniform enzyme concentration using a finite number of
molecules. However, we do not restrict the diffusion of information $\A$ molecules.
If an intermediate $\EA$ molecule reaches the boundary of $\Ve$, then we
probabilistically decompose the molecule using the probabilities for the
unbinding and degradation reactions as calculated by (\ref{AUG12_23}).
As long as $\Ve$ is sufficiently large, such that its boundary is
far away from the receiver, then the impact of these forced
reactions on the observations made at the receiver is negligible.
In our simulations, we ensure that the side length of $\Ve$ is at least three
times greater than $\radmag{0}$, i.e., the distance from the transmitter to
the center of the receiver, and this is sufficient to ignore the
behavior at the boundary of $\Ve$ given the system parameter values that
we use.

\subsection{Selecting Component Parameters}

Most enzymes are proteins and are usually on the order of less than 10\,nm in
diameter; see \cite[Ch. 4]{RefWorks:588}. From
(\ref{JUN12_60}), smaller molecules diffuse faster, so we favor
small molecules as information molecules. Many common small
organic molecules, such as glucose, amino acids, and nucleotides, are about 1\,nm
in diameter. In the limit, single covalent bonds between two atoms are about
0.15\,nm long; see \cite[Ch. 2]{RefWorks:588}.

Higher rate constants imply faster reactions. Bimolecular rate constants
can be no greater than the collision frequency between the two reactants, i.e.,
every collision results in a reaction. The largest possible value of $\kth{1}$
is on the order of $1.66\times 10^{-19}\,\molecule^{-1}\metre^3\,\second^{-1}$;
see \cite[Ch. 10]{RefWorks:585} where the limiting rate is listed as on
the order of $10^8\,$L/mol/s. $\kth{-1}$ and $\kth{2}$ usually
vary between 1 and
$10^5\,\second^{-1}$, with values as high as $10^7\,\second^{-1}$.
In theory, we
are not entirely limited to pre-existing enzyme-substrate pairs; protein and
ribozyme engineering techniques can be used to modify and optimize the enzyme
reaction rate, specificity, or thermal stability, or modify enzyme function in
the presence of solvents; see \cite[Ch. 10]{RefWorks:588}.

\section{Numerical and Simulation Results}
\label{sec_results}

We present simulation results for an environment with
a viscosity of
$10^{-3}\,\textnormal{kg}\cdot\metre^{-1}\second^{-1}$ and temperature of
$25\,^{\circ}\mathrm{C}$.
We compare three sets of system
parameters, as described in Table~\ref{table_accuracy}. 
The chosen values of time step $\Delta t$ are selected large enough
so that the root mean square separation between any two enzyme and
information molecules in a single time step, $\stepl$, is
much greater than the binding radius, $\radbind$
(we include the
values of $\stepl$ and $\radbind$ calculated for each system
in Table~\ref{table_accuracy}), and small enough so that a given information
$\A$ molecule is unlikely to enter and exit the receiver without being observed
(the root mean square displacement of a single $\A$ molecule in a single
time step along one dimension, $\sqrt{2\Dx{\A} t}$, is always less than $\stepl$,
and $\stepl$ is about half of the radius of the receiver).
System 3 is identical to
System 1 except for a larger value of $\Nemit$. $\Ne$ is chosen so that the
enzyme concentration in all systems is equivalent to $166\,\mu\textnormal{M}$
(i.e., micromolar),
which is high for a cellular enzyme; see \cite{RefWorks:632}.
In comparison to the
limiting values of reaction rate constants discussed in
the previous section, the reaction rate constants
$\kth{1}$ and $\kth{2}$ for Systems 1 and 3 are
relatively high due to the small size of the environments.
The numbers of molecules
$\Nemit$ and $\Ne$ and the size of the environments
are kept deliberately low in order to ease computation time.

\ifOneCol
\else
	\tableAccuracy{!tb}
\fi

\subsection{Accuracy of Expected Number of Molecules}
In Fig.~\ref{fig_accuracy}, we compare the observed number of
molecules for Systems 1 and 2 due to a single transmission.
The observed number of $\A$ molecules via simulation is averaged over at least
6000 independent emissions of $\Nemit$ molecules by the transmitter at $t = 0$.
We measure the number of information molecules observed over time, in comparison to the
lower bound expression (\ref{JUN12_47_mult}) and the expected number without enzymes
in the environment as given by (\ref{JUN12_47_mult}) for $\Cx{\Etot} = 0$.

\ifOneCol
\else
	\figAccuracy{!tb}
\fi

We clearly see in Fig.~\ref{fig_accuracy} that the receivers in Systems 1 and 2
have the same lower bound on $\Nobsavg{t}$, the expected number of observed
information molecules, when we account for System 2's longer diffusion time as
its receiver is placed further away.
The simulated number of observed molecules of both
systems over time is close to the derived lower bound curve; the lower bound
expression is accurate for describing the reduction in $\Nobsavg{t}$
in comparison to not having enzymes present. However, the
simulated value of $\Nobsavg{t}$ for System 2 is visibly closer to the
analytical lower bound expression than that for System 1.
Further study
of the effect of the environmental parameters (including chemical reactivity
and the number of molecules) on the accuracy of the analytical lower bound
expression can be found in \cite{RefWorks:706}.

For most of the remaining results, we focus on System 1 because it has a less
accurate lower bound on the expected number of observed molecules and also
because it has overall fewer molecules in the environment
so that its simulations can
be executed more efficiently.

\subsection{Selection of Bit Interval}

From (\ref{JUN12_68}) and (\ref{JUN12_73}) (or by observation of
Fig.~\ref{fig_accuracy}), we calculate that the maximum number of expected
molecules for System 1 should be observed at times $\tmax = 25.68\,\mu\second$
and $\tmax\big|_{\Cx{\Etot} = 0} = 34.36\,\mu\second$ with and without active
enzymes, respectively. At these times, the expected number of observed
molecules is $\Nobsavgmax = 2.92$ and $\Nobsavgmax\big|_{\Cx{\Etot} = 0} =
5.20$, respectively. We are interested in solving (\ref{JUN12_75}) to
get a sense of how long we should wait after an emission from the transmitter
before sending another bit.

In Fig.~\ref{fig_interval}, we solve (\ref{JUN12_75})
for the cases of enzymes present and absent by using upper bounds
(\ref{JUN12_77}) and (\ref{JUN12_80}), respectively. We also solve
(\ref{JUN12_75}) numerically. We see that the bound (\ref{JUN12_77}),
for enzymes present, is quite accurate if the fraction of molecules
expected at the end of the interval is between $30\%$ and $80\%$ of
the expected maximum (representing between 1 and 2.3 molecules expected on
average), whereas the bound (\ref{JUN12_80}), for enzymes absent, improves with
time as fewer molecules are expected.

\ifOneCol
\else
	\figInterval{!tb}
\fi

Whether comparing the bounds or the numerical solutions,
Fig.~\ref{fig_interval} shows that the transmitter can emit much more
frequently with less risk of \ISI\, if enzymes are present. For example, we may
desire to have no more than $30\%$ of $\Nobsavgmax$ within the
receiver at the end of the bit interval $\T$.
Using the numerical solution, we see that we would need to wait about
$170\,\mu\second$ if there were no enzymes present, but only about
$70\,\mu\second$ with enzymes present. This result suggests that we can increase
the data rate by about $150\%$ with a comparable
level of relative \ISI. For lower levels of \ISI\,, the numerical solutions
suggest even higher increases in data transmission.
We emphasize that solving (\ref{JUN12_75}) is insufficient in
evaluating the \ISI\, for a given set of system parameters, but it allows us to
get a sense of what an appropriate $\T$ might be.

\subsection{Detection Probability of One Bit Interval}

Before we consider the bit error rate over a lengthy data transmission, we
consider the detection probability for the first emission by the transmitter.
This enables us to focus on evaluating the accuracy of our expressions
derived for the probability of the observed number of molecules being equal to
or above some threshold; namely, we consider the binomial distribution
(\ref{JUN12_81}), which is exact for a given $\Pobsx{t}$ (recall that we have
(\ref{AUG12_59_not_DMLS}), a lower bound on $\Pobsx{t}$),
and the Poisson and
Gaussian approximations (\ref{JUN12_81_Poiss}) and (\ref{JUN12_90}),
respectively. We evaluate the detection probabilities at time $\tmax
\approx25.5\,\mu\second$ since we have to make observations at multiples of
$\Delta t$ ($\Delta t = 0.5\,\mu\second$ for System 1),
and we compare with the number of $\A$ molecules observed via
simulation as averaged over 6000 independent emissions by the transmitter at $t
= 0$. The results are presented in Fig.~\ref{fig_single} for $1 \le \thresh \le
5$.

\ifOneCol
\else
	\figSingle{!tb}
\fi

In Fig.~\ref{fig_single}, we see that the detection probability can be kept
above $0.5$ for $\thresh \le 3$, which we expect since $\Nobsavgmax = 2.92$.
The binomial distribution based on $\Pobsx{t}$ from (\ref{AUG12_59_not_DMLS})
returns detection probabilities that are comparable to those found via
simulation; for $\thresh = 2$, the detection probability found via simulation
and via the binomial distribution are both about $0.8$. We also see that the
Poisson approximation is indistinguishable from the binomial distribution,
whereas the Gaussian approximation has a notable loss in accuracy.

\subsection{Bit Error Rate of Multiple Intervals}

We now assess the bit error probability for System 1 transmitting multiple bits
by comparing the evaluation of (\ref{SEP12_07}) with simulation
results. We choose the Poisson approximation for evaluating the
expected $\Pe{j}$ because of its high accuracy to the binomial distribution for
System 1. We also select either $\T = 50\,\mu\second$ or $\T = 120\,\mu\second$.
For $\T = 120\,\mu\second$, we see from Fig.~\ref{fig_interval} that the
expected number of molecules at the end of a bit interval due to a single
emission of molecules is less than $20\%$ of the maximum when enzymes are
present but more than $40\%$ of the maximum when there are no enzymes. For $\T =
50\,\mu\second$, the expected \ISI\, is even higher.

First, we consider a known data sequence in order to compare the
accuracy of missed detection (incorrectly detecting a 1 instead of a 0)
versus false alarm (incorrectly detecting a 0 instead of a 1).
The transmitter emits molecules according to a sequence of five consecutive
$1$s followed by five consecutive $0$s. In Fig.~\ref{fig_known_data}, we track
the receiver error probability $\Pe{j}$ over time for $\T = 120\,\mu\second$,
where the simulation results are averaged over 35000 independent transmissions.
Receiver errors within the first five bit intervals are missed
detections, whereas errors within the last five bit intervals are false alarms.
We evaluate the error probability using only knowledge of the current bit,
using knowledge of the current and previous bits,
and using knowledge of the current and all previous bits (up to
nine). We set decision threshold $\thresh = 1$ as we will later see that it is
the optimal threshold for System 1 when $\T = 120\,\mu\second$.

\ifOneCol
\else
	\figKnown{!tb}
\fi

In Fig.~\ref{fig_known_data}, we see via both simulation and evaluation of
(\ref{SEP12_07}) that the error probability reaches a floor on missed detection
with repeated $1$s and tends to zero on false alarm with repeated $0$s, which is
an intuitive result. Note that, when the transmitted bit changes at interval
$6$, the error probability assuming no \ISI\, immediately drops to zero whereas
all other curves spike sharply upwards, showing
a high probability of false alarm when a $0$ is
transmitted after a $1$. The error probability using knowledge of only the
current and previous bits drops to zero by interval $7$, even though the error
measured via simulation and evaluated by considering the current and all
previous bits shows a non-negligible error probability of about $1\%$. All
evaluations of (\ref{SEP12_07}) appear to over-estimate missed detection and
under-estimate false alarm; this makes sense since the underlying probability
of observing an information molecule is a lower bound. Thus, the
accuracy of the expected error probability, even when considering all previous
bits, becomes quite poor when consecutive zeros are transmitted. However, we
will next see that this does not have a noticeable effect on the average bit
error probability for a random transmission in System 1.

We now assess the mean receiver error probability, $\Peavg$, as a function
of the bit decision threshold where we generate a long random source
transmission ($50$ bits). We assume no \emph{a priori} knowledge of
the transmitted data when calculating the expected error probability $\Pe{j}$
from (\ref{SEP12_07}), where all prior bit intervals
are considered and we average $\Pe{j}$ over 1000 random bit sequences,
and $\Peavg$ is evaluated by averaging $\Pe{j}$ over
all $j$.
The results are presented in Fig.~\ref{fig_error_vs_thresh} for
$\T=50\,\mu\second$ and $\T=120\,\mu\second$ where we set
the \emph{a priori} bit probabilities $\Pone = \Pzero =
0.5$. We also consider $\T=120\,\mu\second$ when there are no enzymes present.
Simulation results are
averaged over 6000 independent transmissions.

\ifOneCol
\else
	\figError{!tb}
\fi

We see that in Fig.~\ref{fig_error_vs_thresh}
the optimal decision threshold for System 1 and $\T = 120\,\mu\second$ is
$\thresh = 1$ with enzymes and $\thresh = 5$ without enzymes, whereas the
optimal threshold when $\T = 50\,\mu\second$ is $\thresh = 2$. These differences
make intuitive sense; when the bit interval is shorter or enzymes are
not present, there is more \ISI\, from previous bits such that a lower decision
threshold can result in many more false alarms. The minimum error probability
is much lower for $\T = 120\,\mu\second$ with enzymes; just over $0.05$ versus
over $0.12$ for $\T = 120\,\mu\second$ without
enzymes and for $\T = 50\,\mu\second$.

The error
expected by the evaluation of (\ref{SEP12_07}) with enzymes is much more
accurate than what we might expect from Fig.~\ref{fig_known_data} alone;
a long sequence of consecutive zeros is unlikely in a random transmission, and
the slight over-estimation of missed detection is on averge balanced by the
under-estimation of false alarm.
The primary observation in Fig.~\ref{fig_error_vs_thresh} is that, by adding
enzymes, the data transmission rate can be significantly increased (more than
doubled here) while maintaining the same expected error probability, or the
bit error probability can be significantly improved
for the same data transmission
rate.

Finally, we note that the error probabilities presented in this section are not
very low in the context of information transmission. By deliberately selecting a
system with a low number of molecules, we were limited by a low
expected maximum number of information molecules $\Nobsavgmax$. For contrast, we
consider System 3, for which we can expect $\Nobsavgmax =
11.69$ molecules to be observed at $\tmax = 25.68\,\mu\second$ when there is a
single emission. We evaluate
the average error probability $\Peavg$ when the transmitter in
System 3 emits a stream of $50$ bits with $\T = 120\,\mu\second$.
The results
are plotted in Fig.~\ref{fig_error_vs_thresh_system2}, where we see that for the
optimal threshold $\thresh = 4$, the expected error probability is about
$1.5\times10^{-3}$, much less than those observed for System 1, and the observed
error probability is about $10^{-3}$. The larger deviation between expected and
simulated results for System 3 relative to System 1 is because the lower bound
expression on the PDF (\ref{AUG12_59_not_DMLS}) is not as tight for
System 3; the over-estimation of missed detection and the under-estimation of
false alarm are more evident for System 3 than they are for System 1 as
shown in Fig.~\ref{fig_error_vs_thresh}.

\ifOneCol
\else
	\figErrorTwo{!tb}
\fi

\section{Conclusions and Future Work}
\label{sec_concl}

In this paper, we expanded upon the physical system model that we developed in
\cite{RefWorks:631} for the transmission of impulses
of molecules being released into a
propagation environment that contains diffusing enzymes. We derived a
lower bound expression on the expected number of information molecules at
the receiver. We showed how the expected signal degradation can be used
to predict an appropriate bit interval length.
We then derived the expected error rate for a simple receiver
scheme as a function of the current and all previous emissions. Our
results showed that the expected probability of error can be accurately
represented by the Poisson approximation and agrees with the error
probabilities observed via simulation. The presence of enzymes
was shown to enable a decrease in the probability of error or an increase
in the data transmission rate.

Our on-going work includes considering
the impact of flow, external noise sources,
and multiuser interference on a diffusive communication link.
We are studying the design of practical diffusive detectors and comparing
their performance to that of the optimal detector. We are
also formalizing tractable optimization problems to
minimize the probability of error when selecting the bit decision
threshold and other transmission parameters.

\bibliography{../references/nano_ref}

\newpage

\begin{IEEEbiography}[{\includegraphics[width=1in,height=1.25in,
clip,keepaspectratio]{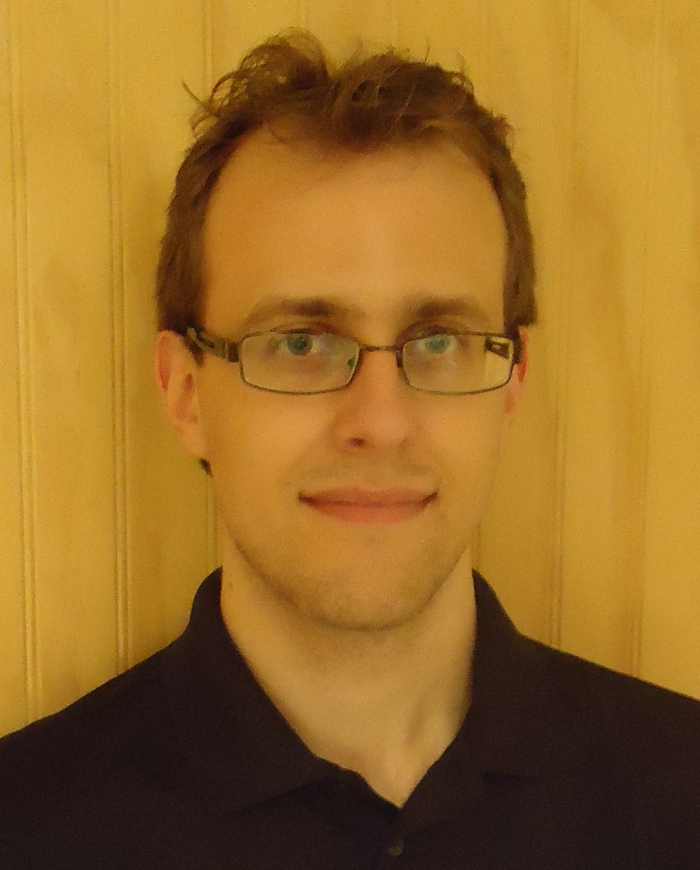}}]{Adam Noel}
(S'09) received the B.Eng. degree from Memorial University in 2009 and the
M.A.Sc.
degree from the University of British Columbia (UBC) in 2011, both in
electrical engineering.
He is now a Ph.D. candidate in electrical
engineering at UBC, and in 2013 was a visiting researcher
at the Institute for Digital Communications,
Friedrich-Alexander-Universit\"{a}t Erlangen-N\"{u}rnberg. His research
interests include wireless communications and how traditional communication
theory applies to molecular communication.
\end{IEEEbiography}

\vspace*{-10cm}

\begin{IEEEbiography}[{\includegraphics[width=1in,height=1.25in,
clip,keepaspectratio]{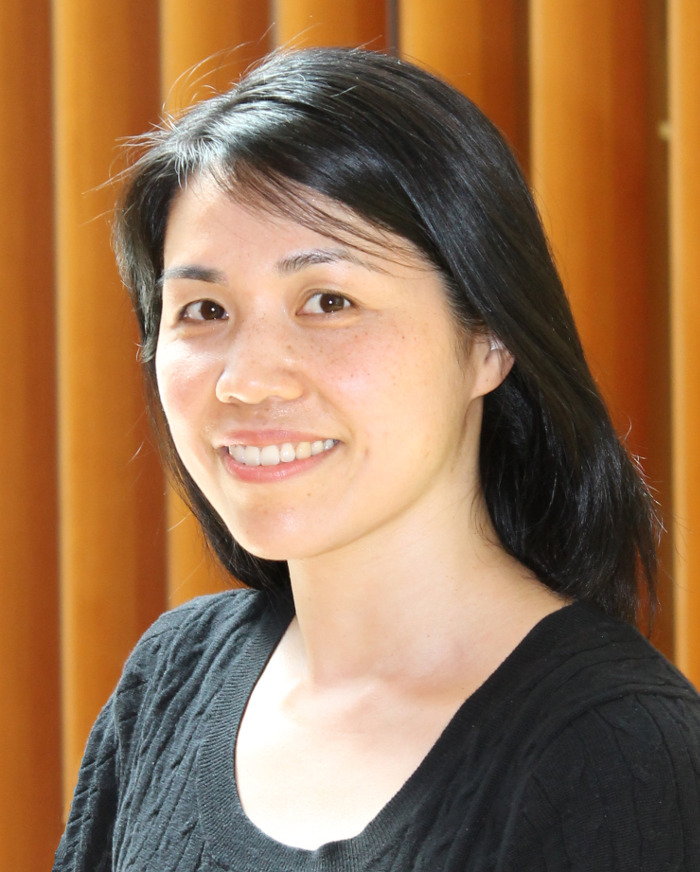}}]{Karen C. Cheung}
received the B.S. and Ph.D. degrees in bioengineering from the University of
California, Berkeley, in 1998 and 2002, respectively. From 2002 to 2005, she was
a postdoctoral researcher at the Ecole Polytechnique Fédérale de Lausanne,
Lausanne, Switzerland. She is now at the University of British Columbia,
Vancouver, BC, Canada. Her research interests include lab-on-a-chip systems for
cell culture and characterization, inkjet printing for tissue engineering, and
implantable neural interfaces.
\end{IEEEbiography}

\vspace*{-10cm}

\begin{IEEEbiography}[{\includegraphics[width=1in,height=1.25in,
clip,keepaspectratio]{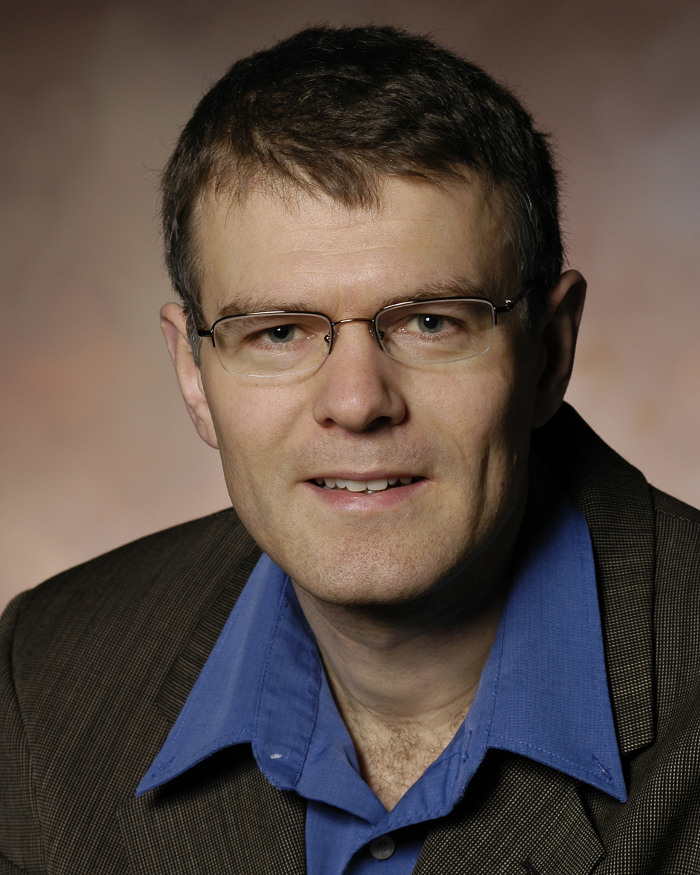}}]{Robert Schober}
(S'98, M'01, SM'08, F'10)
received the Diplom (Univ.) and the Ph.D. degrees in electrical engineering from
the University of Erlangen-Nuremberg in 1997 and 2000, respectively.
Since May 2002 he has been with the University of British Columbia (UBC),
Vancouver, Canada, where he is now a Full Professor.
Since January 2012 he is an Alexander
von Humboldt Professor and the Chair for Digital Communication at the Friedrich
Alexander University (FAU), Erlangen, Germany. His research interests fall into
the broad areas of Communication Theory, Wireless Communications, and
Statistical Signal Processing.
He is currently the
Editor-in-Chief of the IEEE Transactions on Communications.

\end{IEEEbiography}

\ifOneCol
	\newpage
	
		\figEnz{H}
		
	\newpage
	
		\figVenz{H}
	
	\newpage
	
		\tableAccuracy{H}
	
	\newpage
	
		\figAccuracy{H}
	
	\newpage
	
		\figInterval{H}
	
	\newpage
	
		\figSingle{H}
	
	\newpage
	
		\figKnown{H}
	
	\newpage
	
		\figError{H}
	
	\newpage
	
		\figErrorTwo{H}
	
	\newpage
	
	\section*{Figure captions}
	
	Fig. 1 \quad A sample comparison of the expected concentration of information
		molecules at a receiver with and without enzymes present in the propagation
		environment. In each case, the transmitter emits two impulses of molecules.
		The relative quantity of \ISI\, from the first impulse, shown as a thicker
		red line, is much greater without active enzyme molecules.
	
	Fig. 2 \quad The bounded space $\Ve$ showing a uniform
		distribution of enzyme $\En$ molecules
		(enzymes are shown as circles with vertical
		lines through them). $\Ve$ inhibits the passage of $\En$ so that the
		total concentration of free and bound $\En$ remains constant. Information
		$\A$ molecules (shown as red circles) are emitted by the transmitter and can
		diffuse beyond $\Ve$. Intermediate $\EA$ molecules can form when an $\A$
		molecule binds to an $\En$ molecule. When an intermediate dissociates,
		it can leave the $\A$ molecule degraded
		(shown as a circle with an X through it).
	
	Table 1 \quad System parameters used for numerical and simulation results.
	The values for $\stepl$ and $\radbind$ are calculated from (\ref{AUG12_25})
	and (\ref{AUG12_26}), respectively.
	
	Fig. 3 \quad Assessing the accuracy of the lower bound on the expected
	number of observed information molecules for Systems 1 (above) and 2 (below).
	The two systems have the same lower bound on the expected number of
	observed molecules when we account for System 2's longer diffusion time (the
	receiver is placed further away), but this bound is more accurate for System 2.
	
	Fig. 4 \quad Solving (\ref{JUN12_75}) for System 1 to determine how long it
	would take after a transmitter's single emission for the expected number of
	information molecules to decay to threshold fraction $\threshInterval$. The
	inequality is solved both numerically and by using upper bounds
	(\ref{JUN12_77}) and (\ref{JUN12_80}) for System 1 having enzymes
	present and absent, respectively.
	
	Fig. 5 \quad Evaluating the detection probability for the first bit in System 1,
	i.e., $\Pr(\dataObs{1} = 1 | \data{1} = 1)$, as a function of decision
	threshold $\thresh$.
	
	Fig. 6 \quad Evaluating the error probability of System 1 over time with
	bit interval $\T =
	120\,\mu\second$ and a known transmission sequence; five $1$s followed by five
	$0$s.
	
	Fig. 7 \quad Evaluating the error probability of System 1 as a function of the bit
	decision threshold $\thresh$ at the receiver for bit interval
	$\T=50\,\mu\second$ and
	$\T=120\,\mu\second$ with enzymes and $\T=120\,\mu\second$ without enzymes. The
	transmission is a sequence of $50$ randomly generated bits.
	
	Fig. 8 \quad Evaluating the error probability of System 3 as a function of
	the bit decision threshold $\thresh$ at the receiver for
	bit interval $\T=120\,\mu\second$.
\fi

\end{document}